\documentclass[prd,floatfix,showpacs]{revtex4}
\usepackage{epsfig}

\newcommand{\eq}[1]{Eq.~(\ref{#1})}

\newcommand{\be}{\begin{equation}}
\newcommand{\ee}{\end{equation}}
\newcommand{\bea}{\begin{eqnarray}}
\newcommand{\eea}{\end{eqnarray}}
\newcommand{\ben}{\begin{eqnarray*}}
\newcommand{\een}{\end{eqnarray*}}

\newcommand{\DS}{Dyson--Schwinger }
\newcommand{\BS}{Bethe--Salpeter }
\newcommand{\ST}{Slavnov--Taylor }
\newcommand{\YM}{Yang--Mills }

\newcommand{\w}{\omega}
\newcommand{\e}{\varepsilon}
\newcommand{\al}{\alpha}
\newcommand{\ba}{\beta}
\newcommand{\ga}{\gamma}
\newcommand{\G}{\Gamma}
\newcommand{\de}{\delta}

\newcommand{\si}{\sigma}

\newcommand{\ro}{\rho}
\newcommand{\la}{\lambda}

\newcommand{\ka}{\kappa}
\newcommand{\ta}{\tau}
\newcommand{\et}{\eta}

\renewcommand{\th}{\theta}

\renewcommand{\div}{\vec{\nabla}}

\newcommand{\s}[2]{{#1}\!\cdot\!{#2}}
\newcommand{\ov}[1]{\overline{#1}}
\newcommand{\dk}[1]
{\,\,\,\raisebox{-0.4ex}{\large $\bar{}$}\!\!d\,{#1}\,}

\newcommand{\ev}[1]{<\!\!{#1}\!\!>}

\begin{document}

\title{Coulomb gauge confinement in the heavy quark limit}
\author{C.~Popovici}
\author{P.~Watson}
\author{H.~Reinhardt}
\affiliation{Institut f\"ur Theoretische Physik, Universit\"at
T\"ubingen, Auf der Morgenstelle 14, D-72076 T\"ubingen, 
Deutschland}
%\date{\today}
\begin{abstract}
The relationship between the nonperturbative Green's functions of 
Yang-Mills theory and the confinement potential is investigated. 
By rewriting the generating functional of quantum chromodynamics 
in terms of a heavy quark mass expansion in Coulomb gauge, 
restricting to leading order in this expansion and considering 
only the two-point functions of the Yang-Mills sector, the 
rainbow-ladder approximation to the gap and Bethe-Salpeter 
equations is shown to be exact in this case and an analytic, 
nonperturbative solution is presented.  It is found that there 
is a direct connection between the string tension and the temporal 
gluon propagator.  Further, it is shown that for the 4-point quark 
correlation functions, only confined bound states of color-singlet
quark-antiquark (meson) and quark-quark (baryon) pairs exist.
\end{abstract}
\maketitle

\section{Introduction}
\setcounter{equation}{0}
In quantum chromodynamics [QCD] there are two primary aspects of
the theory that remain to this day elusive: confinement and 
dynamical chiral symmetry breaking.  Both involve the existence of
a nontrivial, nonperturbative scale that is generated dynamically. 
Typically when one works with a physical theory such as QCD, one 
does not work directly with physical quantities, but rather with 
(the more readily calculable) gauge-fixed Green's functions.  
In the process of `solving' the theory, one must regularize and 
renormalize at some scale, assigning the various parameters of the
theory physical values such that the dynamically generated 
nonperturbative scale acquires physical meaning.  This naturally 
raises the question: is there a simple relationship between the 
Green's functions and the nonperturbative physical scale 
determined externally?

Further to this, Yang--Mills theory as the sub-theory of QCD
without quarks does not have any directly observable quantities
(the only observables are the glueballs and their spectrum, 
neither of which can be detected without having a detector made of 
matter, i.e., quarks and electrons).  Yang-Mills theory is known 
to be confining in the sense that Wilson loops in lattice 
simulations exhibit an area law at large separations and this area 
law gives rise to a linearly rising potential, the coefficient of 
which is the string tension.  Now whilst the lattice string tension
exists, it has to be related to some physically observable scale
in the real world and this has been explicitly done
\cite{Sommer:1993ce}.  Unfortunately, the Wilson loop is not
particularly amenable to calculation in continuum Yang--Mills
theory (for recent work on this subject, see Ref.~\cite{Pak:2009em}
and references therein), although it is the most natural quantity 
on the lattice.  However, the physical interpretation of the Wilson
loop of pure Yang--Mills theory as corresponding to two infinitely
heavy quark sources connected via gluon exchange may provide a way 
of sidestepping the difficulties of considering the (gauge 
invariant) Wilson loop in the continuum and providing the 
connection between the (gauge-fixed) Green's functions of \YM 
theory and physical quark confinement.  The heavy quark mass 
expansion (and the effective theory associated with it) is a well
established area of QCD and indeed uses phenomenologically 
motivated potentials in place of the Yang--Mills sector 
\cite{Neubert:1993mb, Mannel:1992fx,Grinstein:1991ap} (see
also, for example, Ref.~\cite{Eichten:1980mw}).  Moreover, in the
heavy quark mass limit the spin of the quark decouples from the 
system (the so-called heavy quark symmetry) leading to dramatic
simplifications.

In full QCD, the Bethe--Salpeter equation is one tool for 
describing systems composed of quark pairs with interactions given
via the Green's functions of the theory. In the light quark sector
where one is concerned primarily with dynamical chiral symmetry
breaking and the spectrum of the light mesons, covariant gauge \BS
studies have proven to be extremely useful for phenomenology 
(e.g., Refs.~\cite{Maris:1999nt,Alkofer:2002bp} or the 
contemporary review Ref.~\cite{Fischer:2006ub} and references 
therein).  Such studies are typically restricted to the 
rainbow--ladder truncation, where the \BS kernel is reduced to the
single exchange of a dressed gluon and so that the problem is 
tractable.  However, it is known that the \BS kernel is not so 
simple and indeed, there has been recently some considerable 
attention focused on going beyond the rainbow--ladder
approximation both in terms of vertex corrections (see for example
Refs.~\cite{Williams:2009wx,Watson:2004kd,Matevosyan:2006bk,
Bhagwat:2004hn,Bender:2002as,Bender:1996bb}) and in unquenching 
effects (e.g., Refs.~\cite{Williams:2009wx,Fischer:2009jm,
Fischer:2005en,Watson:2004jq}).  Whilst more sophisticated kernels
do entail a considerable increase in effort, one of the more 
intriguing results of these studies is that (aside from the meson
decay mechanism induced by unquenching \cite{Watson:2004jq})
the rainbow--ladder truncation seems surprisingly robust.  
In Coulomb gauge, there also exist studies of dynamical chiral
symmetry breaking and the \BS equation (see for example Ref.
~\cite{Adler:1984ri} or the more recent Ref.~\cite{Alkofer:2005ug}
and references therein).  Here, potential models (corresponding
to the rainbow--ladder truncation) have been used.

As was recognized early on, Coulomb gauge is an ideal choice for
studying nonperturbative phenomena \cite{Abers:1973qs}, but because
of the inherent noncovariance, significant technical progress has 
been only recently achieved (mainly in Yang--Mills theory). The
Gribov--Zwanziger confinement scenario
\cite{Gribov:1977wm,Zwanziger:1995cv,Zwanziger:1998ez} is 
particularly relevant in Coulomb gauge. In this scenario,
the temporal component of the gluon propagator is infrared
enhanced (providing a long-range confining force) whereas the
transverse spatial components are infrared suppressed (and 
therefore do not appear as asymptotic states).  The relevance
of Coulomb gauge stems from the fact that the system reduces 
naturally to physical degrees of freedom \cite{Zwanziger:1998ez}. 
Using functional methods, it has been formally shown that there 
exists a conserved and vanishing total color charge (along with
the absence of the infamous Coulomb gauge energy
divergences) \cite{Reinhardt:2008pr}.  In addition, the \DS 
equations have been explicitly derived
\cite{Watson:2006yq,Watson:2007vc,Popovici:2008ty}, along with the
\ST identities (for \YM theory) \cite{Watson:2008fb} and one-loop
perturbative results have been calculated
\cite{Watson:2007mz,Watson:2007vc,Popovici:2008ty}. On the lattice,
there are also initial results for the Yang--Mills propagators now
available (although at present the lattice sizes are still modest)
\cite{Burgio:2008jr,Quandt:2008zj}.  Of note is that on the 
lattice, the temporal gluon propagator appears largely independent
of the energy (in the noncovariant Coulomb gauge, the propagators 
are in general dependent on both the energy and spatial momentum 
as separate variables, unlike in covariant gauges) and is 
consistent with a $1/\vec{q}^4$-behavior in the infrared.  
Furhermore in Coulomb gauge the static spatial lattice gluon 
propagator is found to vanish in the infrared and to agree with
Gribov's formula \cite{Burgio:2008jr}. The lattice results support
the findings of the variational approach to continuum \YM theory in
Coulomb gauge \cite{Feuchter:2004mk,Epple:2006hv} (c.f. also
\cite{Epple:2007ut, Szczepaniak:2001rg}).

To establish a connection between the Green's functions of 
Yang--Mills theory and the physical world represented by the 
existence of the string tension that confines quarks, we propose 
to study here the full QCD Bethe--Salpeter equation with a heavy 
quark mass expansion at leading order in Coulomb gauge.  Given
that all quarks are confined irrespective of their mass, it is 
reasonable to suppose that the existence of the confinement 
potential and the value of the string tension are independent of 
the mass and the configuration of the quarks, the only important 
quantity being their separation in this respect.

Now, the heavy quark mass expansion is clearly not a new concept, 
nor is the Bethe--Salpeter equation and the existence of a 
confinement potential.  We shall be working in Coulomb gauge and 
again, there do exist studies of the Bethe--Salpeter equation 
within this context as discussed above.  Thus, we should be 
specific about what is new in this work.  Heavy quark effective 
theory is done typically in covariant gauges.  Separately, in 
Coulomb gauge the study of the Bethe--Salpeter equation has focused
on chiral symmetry breaking and the existence of the confinement 
potential has been taken as being more or less a settled matter 
(see for example Ref.~\cite{Adler:1984ri}).  However, the Coulomb 
gauge Bethe--Salpeter studies have not gone beyond the leading 
(rainbow--ladder) approximation.  In this study, we shall work 
with heavy quarks in Coulomb gauge and with the emergent results 
for the explicit Green's functions of the Yang--Mills sector (at 
least under truncation).  In addition we will be working 
nonperturbatively, deriving and utilizing in part the \ST identity 
for the quark-gluon vertex and its peculiar Coulomb gauge form.  
We thus combine the separate areas of study.  What we shall show 
is that as a result of this combination, the leading 
(rainbow--ladder) approximation is \emph{exact} in the case 
considered, leading directly to the interpretation of quarks being 
confined by a linearly rising potential and providing an explicit 
link between the Green's functions of the nonperturbative 
Yang--Mills sector and the external physical scale.

The paper is organized as follows. In Section~2 we consider the
generating functional of Coulomb gauge QCD at leading order in the 
heavy quark mass expansion. The tree-level quark and antiquark 
propagators are derived and the relevant Feynman prescriptions are 
introduced. In Section~3 the nonperturbative quark gap equation is
discussed. Using the Slavnov-Taylor identities in Coulomb gauge
(derived explicitly in Appendix~\ref{app:stid}) and
truncating the Yang--Mills sector to include only dressed two-point
functions, it is found that the rainbow approximation to the quark 
and antiquark gap equation is nonperturbatively exact. This result
is confirmed by the semiperturbative approach considered in 
Section~4. In Sections~5~\&~6, the Bethe--Salpeter equations for 
quark-antiquark and diquark states are considered. It is shown 
that the  ladder approximation to the \BS kernel is 
nonperturbatively exact and that only color-singlet meson and 
$SU(2)$ baryon states have finite energy solutions. Moreover, a
direct connection between the temporal gluon propagator and the 
string tension is established, at least under the truncation 
scheme considered.

\section{Heavy quark mass expansion}
\setcounter{equation}{0}

In this study, we are concerned with combining nonperturbative 
physics with the heavy quark mass expansion.  The two are not 
automatically compatible; for example, whilst the mass expansion 
assumes that the mass parameter is the largest scale in the 
problem, loop integrals involve momenta up to the UV-cutoff scale 
(supposing for the moment that a UV-cutoff regularization is 
employed) which is also assumed to the the largest scale.  In 
heavy quark effective theory [HQET], the recognition of this 
apparent contradiction leads to the so-called matching procedure 
\cite{Neubert:1993mb}.  This is just one example of the effect of 
the heavy mass expansion and some care must be taken in 
proceeding.  The overriding concern here is that we wish to use 
wherever possible the full nonperturbative QCD functional 
formalism, i.e., the gap and \BS equations and \ST identities for 
the complete quark fields, rather than HQET expressions that refer 
to the heavy quark degrees of freedom.  The effect will be (see 
below) that we are obliged to restrict ourselves to leading order 
in the mass expansion.  This naturally prohibits a quantitative 
analysis involving real quarks (which are not infinitely heavy) 
but does not deflect from the goal of studying the relationship 
between the string tension and the \YM Green's functions.

Throughout this work, we shall use the conventions and notations 
established previously in Refs.~\cite{Popovici:2008ty,
Watson:2008fb,Watson:2007vc,Watson:2006yq}.  We work in Minkowski 
space with the metric $g_{\mu\nu}=\mbox{diag}(1,-\vec{1})$ and all 
minus signs associated with the spatial components of covariant 
and contravariant vectors are explicitly extracted.  Roman 
sub/superscripts ($i$, $j$,\ldots) denote spatial indices, 
superscripts ($a$, $b$,\ldots) denote color indices in the adjoint 
representation.  Configuration space coordinates may be denoted 
with subscript ($x$, $y$,\ldots) when no confusion arises.  Other 
index notation will be explained as the context arises.  The Dirac 
$\ga$-matrices satisfy $\{\ga^\mu,\ga^\nu\}=2g^{\mu\nu}$.  The 
explicit quark contribution to the full QCD generating functional 
is \cite{Popovici:2008ty}
\bea
Z[\ov{\chi},\chi]&=&\int{\cal D}\Phi\exp{\left\{\imath\int d^4x
\ov{q}_\al(x)\left[\imath\ga^0D_0+\imath\s{\vec{\ga}}{\vec{D}}-m
\right]_{\al\ba}q_\ba(x)\right\}}
\nonumber\\&&\times
\exp{\left\{\imath\int d^4x\left[\ov{\chi}_\al(x)q_\al(x)+
\ov{q}_\al(x)\chi_\al(x)\right]+\imath{\cal S}_{YM}\right\}}
\label{eq:genfunc}
\eea
with the temporal and spatial components of the covariant 
derivative (in the fundamental color representation) given by
\bea
D_0&=&\partial_{0}-\imath gT^a\si^a(x),\nonumber\\
\vec{D}&=&\div+\imath gT^a\vec{A}^a(x).
\eea
In the above, ${\cal D}\Phi$ generically denotes the functional 
integration measure over all fields present.  $q_{\al}(x)$ denotes 
the full quark field ($\ov{q}$ is the conjugate or antiquark 
field) where in this case, the subscript $\al$ refers to the 
fundamental color, flavor and spin attributes collectively.  The 
sources $\chi_\al(x)$ and $\ov{\chi}_\al(x)$ are for the complete 
quark fields.  The \YM contribution to the generating functional 
(which will only be of direct relevance to this study in 
Appendix~\ref{app:stid}) is in the standard, second order 
formalism \cite{Watson:2008fb,Watson:2007vc}.  Additionally, 
$\vec{A}$ and $\si$ refer to the spatial and temporal components 
of the gluon field, respectively.  The $T^a$ are the (Hermitian) 
generators of the $SU(N_c)$ group satisfying $[T^a,T^b]=\imath 
f^{abc}T^c$ (with the fully antisymmetric structure constants 
$f^{abc}$) and normalized via $\mbox{Tr}(T^aT^b)=\de^{ab}/2$.  For 
later use we introduce the color factor associated with the quark 
self-energy:
\be
C_F=\frac{N_c^2-1}{2N_c}.
\label{eq:casimir1}
\ee

Now consider the following decomposition of the quark and 
antiquark fields:
\bea
q_\al(x)=e^{-\imath mx_0}\left[h(x)+H(x)\right]_\al,&h_\al(x)=
e^{\imath mx_0}\left[P_+q(x)\right]_\al,&H_\al(x)=e^{\imath mx_0}
\left[P_-q(x)\right]_\al
\nonumber\\
\ov{q}_\al(x)=e^{\imath mx_0}\left[\ov{h}(x)+\ov{H}(x)\right]_\al,
&\ov{h}_\al(x)=e^{-\imath mx_0}\left[\ov{q}(x)P_+\right]_\al,
&\ov{H}_\al(x)=e^{-\imath mx_0}\left[\ov{q}(x)P_-\right]_\al
\label{eq:qdecomp}
\eea
where the (spinor) projection operators are
\be
P_\pm=\frac{1}{2}(\openone\pm\ga^0),\;\;\;\;P_++P_-=\openone,
\;\;\;\;P_+P_-=0,\;\;\;\;P_\pm^2=P_\pm.
\ee
This decomposition is a particular case of the heavy quark 
transform underlying HQET \cite{Neubert:1993mb}.  There, one 
recognizes that a heavy quark within a hadron is almost on-shell 
and moves with the hadron velocity $v$ such that the 4-momentum 
can be written $p^\mu=mv^\mu+k^\mu$ where $|k|\ll m|v|$ and $v^2=1$ 
(such that when $|k|=0$, $p^2=m^2$).  One then uses the general 
projectors $P_\pm=(\openone\pm\slash\!\!\!{v})/2$ and with the 
exponential terms $e^{\pm\imath mv\cdot x}$.   The case used here 
corresponds to the rest frame of the quark, $v^\mu=(1,\vec{0})$, 
but within the context of the generating functional is simply a 
choice of (arbitrary) decomposition that will prove useful in 
Coulomb gauge.  In fact, this choice will result in the 
simplification whereby the spatial components of the \YM Green's 
functions are absent at leading order in the mass expansion.  The 
virtue of the heavy quark decomposition is that the projection 
operators satisfy the following further relations
\be
P_+\ga^0P_+=P_+P_+,\;\;\;\;P_+\ga^0P_-=0,\;\;\;\;P_+\ga^iP_+=0
\ee
such that the following relations hold for the components of the 
quark field:
\be
\ov{h}\ga^0h=\ov{h}h,\;\;\ov{H}\ga^0H=-\ov{H}H,\;\;\ov{h}\ga^0H=
\ov{H}\ga^0h=\ov{h}\ga^ih=\ov{H}\ga^iH=0.
\ee
Inserting the decomposition of the quark fields given by 
\eq{eq:qdecomp} into the generating functional \eq{eq:genfunc} and 
using these relationships one obtains
\bea
Z[\ov{\chi},\chi]&=&\int{\cal D}\Phi\exp{\left\{\imath\int d^4x
\left[
\ov{h}_\al(x)\left[\imath D_0\right]_{\al\ba}h_\ba(x)
\right.\right.}\nonumber\\&&{\left.\left.
+\ov{h}_\al(x)\left[\imath\s{\vec{\ga}}{\vec{D}}\right]_{\al\ba}
H_\ba(x)
+\ov{H}_\al(x)\left[\imath\s{\vec{\ga}}{\vec{D}}\right]_{\al\ba}
h_\ba(x)
+\ov{H}_\al(x)\left[-2m-\imath D_0\right]_{\al\ba}H_\ba(x)\right]
\right\}}
\nonumber\\&&\times
\exp{\left\{\imath\int d^4x\left[e^{-\imath mx_0}\ov{\chi}_\al(x)
\left[h(x)+H(x)\right]_\al+e^{\imath mx_0}\left[\ov{h}(x)+\ov{H}(x)
\right]_\al\chi_\al(x)\right]+\imath{\cal S}_{YM}\right\}}.
\label{eq:genfunc1}
\eea
The difference between the mass expansion (as used here) and HQET 
can now be explained.  Because of our insistence on retaining the 
source terms for the full quark fields, our generating functional 
has only been rewritten in terms of different integration 
variables (note that the Jacobian of the transform is field 
independent and thus trivial) and not altered, although the source 
term expression in the above is slightly modified by the 
appearance of the exponential factors.  We will thus be able to 
work with the full gap and Bethe--Salpeter equations for quarks 
which are derived from $Z[\ov{\chi},\chi]$.  This is in contrast 
to HQET where the quark sources are replaced with sources for the 
$h$-fields (the $H$-fields are integrated out as below).  The 
differences are mainly cosmetic at the level of this study (which 
will concentrate on the leading order in the mass expansion, see 
below) but serve to illustrate some useful points which will be 
highlighted as they arise.

The point of the heavy quark decomposition of the fields is that 
for the $h$-field (`large') components, the quark mass parameter 
$m$ does not appear directly.  When we integrate out the 
$H$-fields however, we get the following expression
\bea
Z[\ov{\chi},\chi]&=&\int{\cal D}\Phi\mbox{Det}\left[\imath D_0+2m
\right]\exp{\left\{\imath\int d^4x\left[
\ov{h}_\al(x)\left[\imath D_0\right]_{\al\ba}h_\ba(x)
\right.\right.}\nonumber\\&&{\left.\left.
+\left[\ov{h}(x)\imath\s{\vec{\ga}}{\vec{D}}+e^{-\imath mx_0}\ov{
\chi}(x)\right]_\al
\left[\imath D_0+2m\right]_{\al\ba}^{-1}
\left[\imath\s{\vec{\ga}}{\vec{D}}h(x)+e^{\imath mx_0}\chi(x)
\right]_\ba\right]\right\}}
\nonumber\\&&\times
\exp{\left\{\imath\int d^4x\left[e^{-\imath mx_0}\ov{\chi}_\al(x)
h_\al(x)+e^{\imath mx_0}\ov{h}_\al(x)\chi_\al(x)\right]+\imath
{\cal S}_{YM}\right\}}.
\label{eq:genfunc2}
\eea
Obviously, since we have integrated out a nontrivial component of 
the original quark field, our expression is nonlocal and this is 
where the heavy mass expansion is necessary.  We notice that 
\cite{Neubert:1993mb}
\bea
\mbox{Det}\left[\imath D_0+2m\right]&\sim&{\cal O}
\left(1/m^2\right),\nonumber\\
\left[\imath D_0+2m\right]_{\al\ba}^{-1}X_\ba&\sim&\frac{1}{2m}
X_\al+{\cal O}\left(1/m^2\right).
\eea
 Leaving the $e^{\pm\imath mx_0}$ factors as they are, we can thus 
write
\bea
Z[\ov{\chi},\chi]&=&\int{\cal D}\Phi\exp{\left\{\imath\int d^4x
\left[
\ov{h}_\al(x)\left[\imath D_0\right]_{\al\ba}h_\ba(x)
+\frac{1}{2m}\left[\ov{h}(x)\imath\s{\vec{\ga}}{\vec{D}}+
e^{-\imath mx_0}\ov{\chi}(x)\right]_\al
\left[\imath\s{\vec{\ga}}{\vec{D}}h(x)+e^{\imath mx_0}
\chi(x)\right]_\al\right]\right\}}
\nonumber\\&&\times
\exp{\left\{\imath\int d^4x\left[e^{-\imath mx_0}\ov{\chi}_\al(x)
h_\al(x)+e^{\imath mx_0}\ov{h}_\al(x)\chi_\al(x)\right]+
\imath{\cal S}_{YM}\right\}}+{\cal O}\left(1/m^2\right).
\label{eq:genfunc3}
\eea
Our generating functional is now local in the fields and arranged 
in an expansion in the parameter $1/m$ (this will be referred to 
as the mass expansion although strictly speaking, it is an 
expansion in the \emph{inverse} mass).  However, locality does not 
mean that the above expression can be directly applied.  Let us 
consider the classical (full) quark field in the presence of 
sources:
\bea
\lefteqn{\frac{1}{Z}\int{\cal D}\Phi q_\al(x)\exp{\left\{\imath
{\cal S}\right\}}=\frac{1}{Z}\frac{\de Z}{\de\imath\ov{\chi}
_\al(x)}}\nonumber\\
&=&\frac{1}{Z}\int{\cal D}\Phi\left\{e^{-\imath mx_0}h_\al(x)+
\frac{e^{-\imath mx_0}}{2m}\left[\imath\s{\vec{\ga}}{\vec{D}}h(x)+
e^{\imath mx_0}\chi(x)\right]_\al\right\}\exp{\left\{\imath
{\cal S}\right\}}+{\cal O}\left(1/m^2\right).
\eea
One sees immediately that even at ${\cal O}(1/m)$, the classical 
quark field has components that involve interaction type terms 
($\vec{D}h\sim\vec{A}h$) brought about by the truncation of the 
nonlocality.  Of course, this is nothing more than the statement 
that the $h$-field is nontrivially (and dynamically) related to 
the full $q$-field.  It also means that if we want to use the 
nonperturbative gap and Bethe--Salpeter equations (i.e., those 
equations derived from the action for the full quark fields $q$ 
and their sources $\chi$ and which are the equations of QCD as 
opposed to HQET) then we cannot expect that the mass expansion can 
realistically be extended far beyond the leading order in order to 
do practical calculations.  As stated previously though, the aim 
is to investigate the connection between the Yang--Mills sector 
and the physical world made of quarks; the string tension that 
represents our goal is not dependent on the quark mass (both light 
and heavy quarks are confined in the same way, as far as we know).
Therefore, we restrict our attention to the leading order in the 
mass expansion as follows (and writing $D_0$ explicitly):
\bea
Z[\ov{\chi},\chi]&=&\int{\cal D}\Phi\exp{\left\{\imath\int d^4x
\ov{h}_\al(x)\left[\imath\partial_{0x}+gT^a\si^a(x)\right]_{\al\ba}
h_\ba(x)\right\}}
\nonumber\\&&\times
\exp{\left\{\imath\int d^4x\left[e^{-\imath mx_0}\ov{\chi}_\al(x)
h_\al(x)+e^{\imath mx_0}\ov{h}_\al(x)\chi_\al(x)\right]+
\imath{\cal S}_{YM}\right\}}+{\cal O}\left(1/m\right).
\label{eq:genfunc4}
\eea

The standard machinery of functional methods is now employed.  
From the observation that up to boundary terms (which are assumed 
to vanish) the integral of a derivative vanishes, we have the 
quark field equation of motion:
\bea
0&=&\int{\cal D}\Phi\frac{\de}{\de\imath\ov{h}_\al(x)}\exp{\left\{
\imath{\cal S}\right\}}\nonumber\\
&=&\int{\cal D}\Phi\left\{\left[\imath\partial_{0x}+gT^a\si^a(x)
\right]_{\al\ba}h_\ba(x)+e^{\imath mx_0}\chi_\al(x)\right\}\exp{
\left\{\imath{\cal S}\right\}}+{\cal O}\left(1/m\right).
\label{eq:feom}
\eea
The field equation of motion for the antiquark gives equivalent 
results and can be neglected.  The generating functional of 
connected Green's functions (for the full quarks) is defined via 
$Z=e^W$.  Denoting the derivatives with respect to sources of 
$W[\ov{\chi},\chi]$ with an obvious bracket notation, the 
classical fields are:
\bea
q_\al(x)=\frac{1}{Z}\int{\cal D}\Phi q_\al(x)\exp{\left\{\imath{
\cal S}\right\}}&=\frac{1}{Z}\frac{\de Z[\ov{\chi},\chi]}{\de\imath
\ov{\chi}_\al(x)}&=:\ev{\imath\ov{\chi}_\al(x)}\nonumber\\
\ov{q}_\al(x)=\frac{1}{Z}\int{\cal D}\Phi \ov{q}_\al(x)\exp{\left\{
\imath{\cal S}\right\}}&=-\frac{1}{Z}\frac{\de Z[\ov{\chi},\chi]}{
\de\imath\chi_\al(x)}&=:-\ev{\imath\chi_\al(x)}.
\eea
Also ($\ro$ is the source for the $\si$-field, implicit within $Z$ 
and $W$),
\be
\int{\cal D}\Phi \si^a(x)q_\al(x)\exp{\left\{\imath{\cal S}
\right\}}=\frac{\de^2Z[\ov{\chi},\chi]}{\de\imath\ro^a(x)\de\imath
\ov{\chi}_\al(x)}=Z[\ov{\chi},\chi]\left[\ev{\imath\ro^a(x)\imath
\ov{\chi}_\al(x)}+\ev{\imath\ro^a(x)}\ev{\imath\ov{\chi}_\al(x)}
\right].
\ee
Notice that from \eq{eq:genfunc4}, the following relations hold:
\be
\frac{\de Z[\ov{\chi},\chi]}{\de\imath\ov{\chi}_\al(x)}=\int
{\cal D}\Phi e^{-\imath mx_0}h_\al(x)\exp{\left\{\imath{\cal S}
\right\}}+{\cal O}\left(1/m\right),\;\;\;\;\frac{\de Z[\ov{\chi},
\chi]}{\de\imath\chi_\al(x)}=-\int{\cal D}\Phi e^{\imath mx_0}
\ov{h}_\al(x)\exp{\left\{\imath{\cal S}\right\}}+{\cal O}\left(1/m
\right)
\ee
(recalling the earlier discussion of neglecting the ${\cal O}
\left(1/m\right)$ terms) such that the field equation of motion,
 \eq{eq:feom}, can be written in terms of derivatives of $W$:
\bea
0&=&\left[\imath\partial_{0x}\right]_{\al\ba}e^{\imath mx_0}\ev
{\imath\ov{\chi}_\ba(x)}+\left[gT^a\right]_{\al\ba}e^{\imath mx_0}
\left[\ev{\imath\ro^a(x)\imath\ov{\chi}_\ba(x)}+\ev{\imath\ro^a(x)}
\ev{\imath\ov{\chi}_\ba(x)}\right]+e^{\imath mx_0}\chi_\al(x)
\nonumber\\&&
+{\cal O}\left(1/m\right).
\eea
Factoring out the exponential terms gives then
\bea
0&=&\left[\imath\partial_{0x}-m\right]_{\al\ba}\ev{\imath\ov{\chi}_
\ba(x)}+\left[gT^a\right]_{\al\ba}\left[\ev{\imath\ro^a(x)\imath
\ov{\chi}_\ba(x)}+\ev{\imath\ro^a(x)}\ev{\imath\ov{\chi}_\ba(x)}
\right]+\chi_\al(x)+{\cal O}\left(1/m\right).\nonumber\\
\label{eq:feom1}
\eea
To continue, we make a Legendre transform in order to construct 
the effective action for the full quark fields.  Explicitly 
separating the quark and Yang--Mills components:
\be
\G[\Phi,\ov{q},q]=W[J,\ov{\chi},\chi]-\imath J_\al\Phi_\al-\imath
\ov{\chi}_\al q_\al-\imath\ov{q}_\al\chi_\al.
\ee
such that
\bea
q_\al(x)=\ev{\imath\ov{\chi}_\al(x)},&&\chi_\al(x)=-\ev{\imath
\ov{q}_\al(x)},\nonumber\\
\ov{q}_\al(x)=-\ev{\imath\chi_\al(x)},&&\ov{\chi}_\al(x)=\ev{
\imath q_\al(x)},\nonumber\\
\si^a(x)=\ev{\imath\ro^a(x)},&&\ro^a(x)=-\ev{\imath\si^a(x)}
\eea
where the same bracket notation for derivatives of the effective 
action with respect to the classical fields is used (there is no 
confusion between the two sets of brackets since the two sets of 
derivatives are never mixed).  $J_\al$ and $\Phi_\al$ denote 
generic gluonic (Yang--Mills) sources and classical fields, 
respectively; the index $\al$ here refers to all attributes of the 
(gluonic) object in question (including its type and position 
argument) and we use the common convention that this index is 
either summed or integrated over as appropriate.  The field 
equation of motion can then be rewritten in terms of derivatives 
of the effective action (which are the proper Green's functions 
when sources are set to zero):
\be
\ev{\imath\ov{q}_\al(x)}=\left[\imath\partial_{0x}-m\right]_
{\al\ba}q_\ba(x)+\left[gT^a\right]_{\al\ba}\left[\ev{\imath\ro^a(x)
\imath\ov{\chi}_\ba(x)}+\si^a(x)q_\ba(x)\right]+{\cal O}\left(1/m
\right).
\label{eq:feom2}
\ee
With the field equation of motion written in the above forms, we 
can now derive the Feynman rules for the quark components of the 
theory (the Yang--Mills parts are already known 
\cite{Watson:2007vc}).  We start with the quark propagator.  
From \eq{eq:feom1}, ignoring interaction terms and functionally 
differentiating we get the tree-level propagator in configuration 
space
\be
0=\left[\imath\partial_{0x}-m\right]_{\al\ba}\ev{\imath\chi_\ga(z)
\imath\ov{\chi}_\ba(x)}^{(0)}-\imath\de_{\ga\al}\de(z-x)+{\cal O}
\left(1/m\right).
\ee
We now write the Fourier transforms to define both the anti- and 
quark propagators, respectively, utilizing the translational 
invariance:
\bea
\ev{\imath\chi_\ga(z)\imath\ov{\chi}_\ba(x)}\equiv W_{q\ov{q}
\ga\ba}(z-x)&=&\int\dk{k}e^{-\imath k\cdot(z-x)}W_{q\ov{q}\ga\ba}
(k),\nonumber\\
\ev{\imath\ov{\chi}_\ba(z)\imath\chi_\al(x)}\equiv W_{\ov{q}q\ba
\al}(z-x)&=&\int\dk{k}e^{-\imath k\cdot(z-x)}W_{\ov{q}q\ba\al}(k)
\eea
such that in momentum space, we get
\bea
0&=&\int\dk{k}e^{-\imath k\cdot(z-x)}\left\{\left[-k_0-m\right]_
{\al\ba}W_{q\ov{q}\ga\ba}^{(0)}(k)-\imath\de_{\ga\al}\right\}+
{\cal O}\left(1/m\right),\label{eq:propmassexp}\nonumber\\
0&=&\int\dk{k}e^{-\imath k\cdot(z-y)}\left\{W_{\ov{q}q\ba\al}^{(0)}
(k)\left[k_0-m\right]_{\al\ga}+\imath\de_{\ba\ga}\right\}+{\cal O}
\left(1/m\right).
\eea
Naively, one would write the solutions as
\be
W_{\ov{q}q\ba\al}^{(0)}(k)=\frac{-\imath\de_{\ba\al}}{\left[k_0-
m\right]}+{\cal O}\left(1/m\right),\;\;\;\;
W_{q\ov{q}\ga\ba}^{(0)}(k)=\frac{-\imath\de_{\ga\ba}}{\left[k_0+
m\right]}+{\cal O}\left(1/m\right).
\ee
However, in order to define the Fourier transform, one must first 
define a prescription for handling the poles in the energy 
integral.  Unlike the conventional tree-level quark propagator, we 
do not have a pair of simple poles in the complex $k_0$-plane, we 
have instead single poles and this is due to the mass expansion.  
For the quark propagator, we write
\be
W_{\ov{q}q\ba\al}^{(0)}(k)=\frac{-\imath\de_{\ba\al}}{\left[k_0-m+
\imath\e\right]}+{\cal O}\left(1/m\right).
\label{eq:quarkpropagator}
\ee
Notice the following.  The tree-level quark propagator is a scalar 
quantity (or rather, diagonal in the outer product of the 
fundamental color, flavor and spinor spaces) as a consequence of 
the mass expansion.  This is one of the most striking features of 
HQET and physically corresponds to the decoupling of the spin from 
the heavy quark system.  $W_{\ov{q}q}^{(0)}$ is identical to the 
heavy quark tree-level propagator \cite{Neubert:1993mb} up to the 
appearance of the mass term.  In HQET, where one uses sources for 
the $h$-fields directly, this term does not appear and is simply 
due to a shift of the energy by an amount $m$ (such that one is 
working with the `small' momenta around the rest mass energy).  We 
retain this term for completeness in order to see what effect it 
may have -- the Yang--Mills sector may not be compatible with such 
a shift in the energy, although this will turn out not to be the 
case.  Also, note that the kinetic term of the tree-level 
propagator would read $-\vec{k}^2/2m$ in the denominator factor 
and is at higher order in the mass expansion.  Such terms are 
obviously important to the UV properties of the loop integrals but 
will not play any role here since we shall be interested in the 
infrared limit.

Returning to the issue of the Feynman prescription, one of the 
direct consequences of the single energy pole is that when one 
considers a closed quark loop (a virtual quark-antiquark pair) at 
lowest order, the energy integral automatically vanishes (see also 
Ref.~\cite{Neubert:1993mb} for an alternate discussion on this 
topic), i.e.,
\be
\int\frac{dk_0}{\left[k_0-m+\imath\e\right]\left[k_0+p_0-m+\imath
\e\right]}=0.
\ee
It can be justified that all such closed quark loops vanish in 
similar fashion (indeed, precisely this type of integral will be 
heavily used in the following Sections); in other words, the heavy 
mass expansion at leading order is quenched.  One interpretation 
of the above result is that due to the presence of only the single 
pole, the heavy quark can only propagate forward in time and the 
corresponding antiquark is not present.  This leads to a 
conundrum: how can we construct a quark-antiquark pair in the \BS 
equation?  The resolution is straightforward -- in the \BS 
equation one is not considering a virtual quark-antiquark pair but 
rather a system composed of two separate unphysical particles.  It 
is known that in Coulomb gauge, Gauss' law forbids the creation of 
a colored state in isolation (the total color charge is conserved 
and vanishing \cite{Reinhardt:2008pr}) and so, the Feynman 
prescription for the quark (or the antiquark) propagator has no 
physical meaning in isolation.  The quenching of the theory means 
that closed quark loops vanish and from covariant gauge \BS 
studies, the connection between such loops and the \BS kernel is 
qualitatively understood \cite{Watson:2004jq}.  In the quenched 
case, the quark and antiquark lines of the \BS equation are never 
connected by a primitive vertex (unlike the closed quark loop).  
Another way of seeing this is to consider the flavor nonsinglet 
\BS equation -- in other words the quark and antiquark 
constituents considered as two distinct flavors.  Because the 
quark and antiquark are genuinely separate (unphysical) particles, 
we are at liberty to choose that the antiquark propagator in the 
\BS equation has the opposite Feynman prescription as follows:
\be
W_{q\ov{q}\ga\ba}^{(0)}(k)=\frac{-\imath\de_{\ga\ba}}{\left[k_0+m+
\imath\e\right]}+{\cal O}\left(1/m\right).
\label{eq:quarkantipropagator}
\ee
As shall be seen, this results in a physical interpretation for 
the quark-antiquark \BS equation as a whole.  Another way of 
viewing the above is to realise that the four components of the 
full quark Dirac equation (quark and antiquark moving either 
forwards or backwards in time according to causality) have been 
separated by the mass expansion (the `large' $h$ or `small' $H$ 
field components) and the breaking of the time reversal symmetry.  
In the context of the mass expansion, there is only the quark or 
antiquark moving forward in time corresponding to the above 
Feynman prescriptions.  The closed quark loop involves a quark 
going backwards in time (similarly for an antiquark loop) which is 
prohibited, or more precisely suppressed by the mass, so that such 
closed quark loop integrals vanish at leading order whereas the 
quark-antiquark \BS equation has a physical solution.

For the proper two-point and three-point functions, we use 
functional derivatives of \eq{eq:feom2} and we get directly in 
momentum space
\bea
\G_{\ov{q}q\al\ba}^{(0)}(k)=\imath\left[k_0-m\right]\de_{\al\ba}+
{\cal O}\left(1/m\right),&&
\G_{q\ov{q}\al\ba}^{(0)}(k)=\imath\left[k_0+m\right]\de_{\al\ba}+
{\cal O}\left(1/m\right),\nonumber\\
\G_{\ov{q}q\si\al\ba}^{(0)a}(k_1,k_2,k_3)=\left[gT^a\right]_
{\al\ba}+{\cal O}\left(1/m\right),&&
\G_{q\ov{q}\si\al\ba}^{(0)a}(k_1,k_2,k_3)=-\left[gT^a\right]_
{\ba\al}+{\cal O}\left(1/m\right).
\label{eq:feyn}
\eea
Importantly, the tree-level spatial quark-gluon vertex does not 
appear at leading order in the mass expansion:
\be
\G_{\ov{q}qA\al\ba i}^{(0)a}=\G_{q\ov{q}A\al\ba i}^{(0)a}={\cal O}
\left(1/m\right).
\ee
Notice the ordering of the indices for the $\G_{q\ov{q}\si}$ 
vertex.  Also notice that $\G_{\ov{q}q\al\ba}^{(0)}(k)=-\G_{q\ov{q}\al
\ba}^{(0)}(-k)$ since the two-point function requires no Feynman 
prescription and is diagonal in the outer product of the 
fundamental color, flavor and spinor spaces.  In addition, $W_{\ov{
q}q}\G_{\ov{q}q}=\openone$ as usual.

Now, because of our insistence of using the full quark sources, 
all the nonperturbative equations involving the quarks (the \DS 
equations for two-point and three-point functions, the \ST 
identities and the \BS equation) will not alter their form at 
leading order in the mass expansion (beyond leading order, as 
previously illustrated, the nontrivial classical field and the 
resolution of the Legendre transform would play a role).  The only 
alterations are that the tree-level factors have been changed and 
we demand that only the leading order terms in the mass expansion 
of the resulting equations are retained.  Obviously, in order to 
solve the nonperturbative system we must further specify our 
truncation scheme.  Recall that at leading order in the mass 
expansion, the theory is already quenched.  The truncation scheme 
we propose is to consider only the dressed two-point functions of 
the Yang--Mills sector (i.e., the nonperturbative gluon 
propagators derived from a hypothetical solution of the complete 
\YM theory).  This amounts to setting all the pure \YM vertices 
and higher $n$-point functions occurring in the quark equations to 
zero.  It is worth pointing out that because of the mass 
expansion, any loop diagram with a tree-level spatial quark-gluon 
vertex will be suppressed, such that the number of loop diagrams 
arising because of the \YM vertices is heavily restricted and 
those loops that do contribute will not include the leading order 
perturbative corrections (the fully temporal gluon Green's 
functions $\G_{\si\si\si},\G_{\si\si\si\si},\ldots$ are zero at 
tree-level).  The most important physical point that will emerge 
is that when we set the \YM vertices to zero, we exclude the 
non-Abelian part of the charge screening mechanism of the quark 
color charge and any potential glueball states.  On the other 
hand, the charge screening mechanism and glueball contributions of 
the gluon field (i.e., the color string) is implicitly encoded in 
the nonperturbative form of the temporal gluon propagator.

With the truncation scheme as outlined above, the Yang--Mills 
sector collapses to the inclusion of a single object: the temporal 
gluon propagator which is written as \cite{Watson:2007vc}
\be
W_{\si\si}^{ab}(k)=\de^{ab}\frac{\imath}{\vec{k}^2}D_{\si\si}(\vec
{k}^2).
\label{eq:gtemp}
\ee
There are three important features to this propagator.  Firstly, 
there are indications on the lattice that the dressing function 
$D_{\si\si}$ is largely independent of energy \cite{Quandt:2008zj}, 
justifying the energy independence of the above form.  Indeed on 
general grounds, the temporal gluon propagator must have some part 
that is constant in the energy in order to cancel closed ghost 
loops and resolve the Coulomb gauge energy divergences (see also 
the more formal considerations of Ref.~\cite{Cucchieri:2000hv}).  
Second, the lattice analysis indicates that the dressing function 
$D_{\si\si}$ is infrared divergent and is likely to behave as $1/
\vec{k}^2$ for vanishing $\vec{k}^2$.  We are interested mainly in 
the relationship between $D_{\si\si}$ (as the input of the 
Yang--Mills sector) and the string tension so we will not need the 
specific form until towards the end.  Third, the product $g^2D_{\si
\si}$ is a renormalization group invariant quantity and thus a 
natural candidate for being relevant to the physical string 
tension \cite{Zwanziger:1998ez,Watson:2008fb}.

\section{Gap equation: nonperturbative treatment}
\setcounter{equation}{0}

Let us begin by considering the \DS equation for the quark 
two-point proper function (the gap equation).  In full QCD (i.e., 
second order formalism, Coulomb gauge without the mass expansion 
and derived from the first order formalism results of Ref.~\cite
{Popovici:2008ty}), it reads ($\dk{\w}=d^4\w/(2\pi)^4$):
\bea
\G_{\ov{q}q\al\de}(k)&=\G_{\ov{q}q\al\de}^{(0)}(k)+\int\dk{\w}&
\left\{\G_{\ov{q}q\si\al\ba}^{(0)a}(k,-\w,\w-k)W_{\ov{q}q\ba\ga}
(\w)\G_{\ov{q}q\si\ga\de}^{b}(\w,-k,k-\w)W_{\si\si}^{ab}(k-\w)
\right.\nonumber\\&&\left.
+\G_{\ov{q}qA\al\ba i}^{(0)a}(k,-\w,\w-k)W_{\ov{q}q\ba\ga}(\w)\G_{
\ov{q}qA\ga\de j}^{b}(\w,-k,k-\w)W_{AAij}^{ab}(k-\w)
\right\}
\label{eq:gap}
\eea
($W_{AA}$ is the spatial gluon propagator, which will be 
unimportant here).  The full quark-gluon vertices obey the \ST 
identity.  The derivation of this identity follows from the 
invariance of the action under a Gauss-BRST transform 
\cite{Zwanziger:1998ez} that is peculiar to Coulomb gauge and is 
similar to the identities derived for the pure \YM sector 
\cite{Watson:2008fb}.  This derivation is presented in 
Appendix~\ref{app:stid} (and is a technical result of this study 
in its own right).  The identity reads, \eq{eq:stid3}:
\bea
k_3^0\G_{\ov{q}q\si\al\ba}^{d}(k_1,k_2,k_3)&=&\imath\frac{k_{3i}}
{\vec{k}_3^2}\G_{\ov{q}qA\al\ba i}^{a}(k_1,k_2,k_3)\G_{\ov{c}c}^
{ad}(-k_3)\nonumber\\&&
+\G_{\ov{q}q\al\de}(k_1)\left[\tilde{\G}_{\ov{q};\ov{c}cq}^{d}
(k_1+q_0,k_3-q_0;k_2)+\imath gT^d\right]_{\de\ba}\nonumber\\&&
+\left[\tilde{\G}_{q;\ov{c}c\ov{q}}^{d}(k_2+q_0,k_3-q_0;k_1)-
\imath gT^d\right]_{\al\de}\G_{\ov{q}q\de\ba}(-k_2)
\eea
where $k_1+k_2+k_3=0$, $q_0$ is the arbitrary energy injection 
scale, $\G_{\ov{c}c}$ is the ghost proper two-point function, 
$\tilde{\G}_{\ov{q};\ov{c}cq}$ and $\tilde{\G}_{q;\ov{c}c\ov{q}}$ are 
ghost-quark kernels associated with the Gauss-BRST transform 
(see Appendix for details).

\begin{figure}[t]
\vspace{0.5cm}
\includegraphics[width=0.6\linewidth]{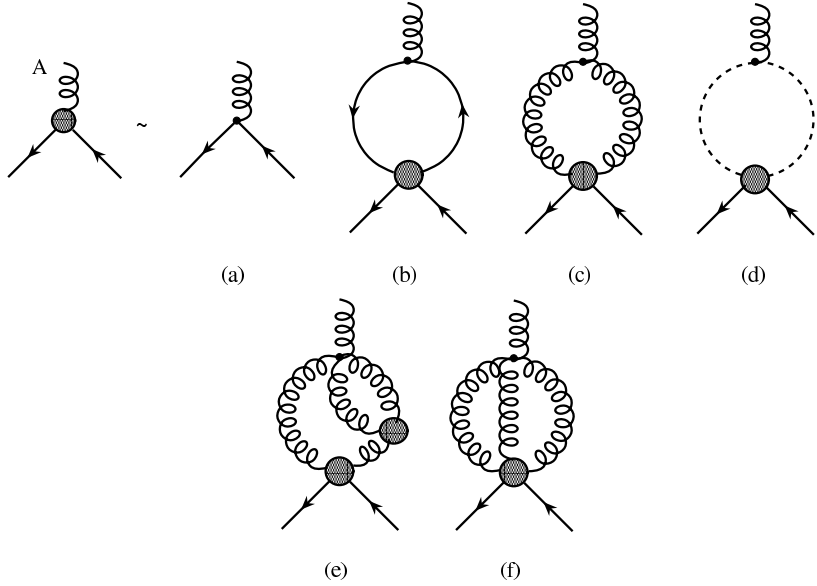}
\caption{\label{fig:qbqa}Diagrams that contribute to the \DS 
equation for the \emph{spatial} quark-gluon vertex (without 
prefactors or signs).  Internal propagators are fully dressed and 
blobs represent dressed proper vertex and (reducible) kernels.  
Internal propagators represented by springs may be either spatial 
($\vec{A}$) or temporal ($\si$) propagators, dashed lines 
represent the ghost propagator and solid lines represent the quark 
propagator.}
\end{figure}

In order to use the \ST identity as input for solving the gap 
equation, we first apply our truncation scheme in the context of 
the heavy mass expansion at leading order.  Starting with the 
dressed spatial quark-gluon vertex, consider the terms that 
contribute to the \DS equation shown schematically in 
Fig.~\ref{fig:qbqa}.  According to the truncation scheme, we set 
all Yang--Mills vertices to zero, meaning that diagrams (c-f) are 
excluded.  This then leaves us with the tree-level term (a) and 
the quark-loop term (b).  However, both of these involve at least 
one tree-level spatial quark-gluon vertex which is not present at 
leading order in the mass expansion.  Thus, we obtain the 
nonperturbative result that
\be
\G_{\ov{q}qA\al\ba i}^{a}(k_1,k_2,k_3)={\cal O}\left(1/m\right).
\label{eq:qbqa}
\ee
It is straightforward to justify that the ghost-quark kernels of 
the \ST identity, given their definition, \eq{eq:kern0}, involve 
\YM vertices and do not contribute here.  Thus, in our truncation 
scheme and at leading order in the mass expansion, the \ST 
identity reads
\be
k_3^0\G_{\ov{q}q\si\al\ba}^{d}(k_1,k_2,k_3)=
\G_{\ov{q}q\al\de}(k_1)\left[\imath gT^d\right]_{\de\ba}-\left[
\imath gT^d\right]_{\al\de}\G_{\ov{q}q\de\ba}(-k_2)+{\cal O}
\left(1/m\right).
\ee
Clearly, the truncation scheme thus results in an Abelian type Ward
identity.
 Moreover, since the temporal quark-gluon vertex is simply 
multiplied by the temporal gluon energy (the primary feature of 
Coulomb, as opposed to covariant, gauge \ST identities) and the 
quark proper two-point function is color diagonal, we can 
immediately write the solution:
\be
\G_{\ov{q}q\si\al\ba}^{d}(k_1,k_2,k_3)=\frac{\imath g}{k_3^0}\left
\{T^d\left[\G_{\ov{q}q}(k_1)-\G_{\ov{q}q}(-k_2)\right]\right\}_{\al
\ba}+{\cal O}\left(1/m\right).
\label{eq:qbqs}
\ee
The above solution is trivially satisfied at tree-level.  
We notice however, that there appears to be a potential problem 
with the energy.  When $k_3^0=0$, but $\vec{k}_3\neq0$ (a 
spacelike gluon configuration), the simple pole must somehow be 
canceled by the difference of proper quark two-point functions.  
Since the spatial momentum configuration is arbitrary, this means 
that $\G_{\ov{q}q}(k)\rightarrow\G_{\ov{q}q}(k_0)+{\cal O}\left(1/m
\right)$.  The demand that the nonperturbative vertex solution to 
the Coulomb gauge \ST identity be free of kinematic divergences 
(here, simply the $1/k_3^0$ factor) is a variation of the familiar 
covariant gauge situation considered in Ref.~\cite{Ball:1980ay}.

Inserting the results, \eq{eq:qbqs} and \eq{eq:qbqa}, for the 
vertices, using the Feynman rules given by \eq{eq:feyn}, with the 
temporal gluon propagator given by \eq{eq:gtemp} and resolving the 
color structure the nonperturbative gap equation, \eq{eq:gap} 
under truncation and at leading order in  the mass expansion thus 
reads 
\be
\G_{\ov{q}q\al\de}(k_0)=\imath\left[k_0-m\right]\de_{\al\de}-g^2C_F
\int\frac{\dk{\w}D_{\si\si}(\vec{k}-\vec{\w})}{(k_0-\w_0)(\vec{k}-
\vec{\w})^2}W_{\ov{q}q\al\ba}(\w_0)\left[\G_{\ov{q}q}(\w_0)-\G_{\ov
{q}q}(k_0)\right]_{\ba\de}+{\cal O}\left(1/m\right).
\label{eq:gap1}
\ee
There exists one particularly simple solution to this equation as 
we shall now demonstrate.  It is given by
\be
W_{\ov{q}q\al\ba}(k)=\frac{-\imath\de_{\al\ba}}{\left[k_0-{\cal C}+
\imath\e\right]}+{\cal O}\left(1/m\right),\;\;\;\;\G_{\ov{q}q\al
\ba}(k)=\imath\de_{\al\ba}\left[k_0-{\cal C}\right]+{\cal O}
\left(1/m\right)
\label{eq:gapsol}
\ee
and where the constant ($\dk{\vec{\w}}=d^3\vec{\w}/(2\pi)^3$)
\be
{\cal C}=m+\frac{1}{2}g^2C_F\int\frac{\dk{\vec{\w}}D_{\si\si}(\vec{
\w})}{\vec{\w}^2}+{\cal O}\left(1/m\right).
\label{eq:const}
\ee
Putting the above solution into the \ST identity, we also have 
that for the vertex
\be
\G_{\ov{q}q\si\al\ba}^{d}(k_1,k_2,k_3)=\left[gT^d\right]_{\al\ba}+
{\cal O}\left(1/m\right).
\label{eq:vsol}
\ee
In other words, the dressed temporal quark-gluon vertex is trivial 
and the gap equation reduces to the rainbow truncation.  There is 
a subtlety to the solution, \eq{eq:gapsol}, involving the ordering 
of the limits in the spatial and temporal integrals and potential 
divergences in the constant given by \eq{eq:const}.  Here, let us 
consider the case when we perform the temporal integral first 
under the condition that the spatial integral is somehow 
regularized and finite (this will be  done throughout the rest of 
the paper).  Inserting the solution given by \eq{eq:gapsol}, the 
gap equation under truncation, \eq{eq:gap1}, and with a 
regularized (i.e., finite) spatial integral denoted by the 
subscript $r$, reads
\bea
{\cal C}_r&=&m+g^2C_F\int_r\frac{\dk{\vec{\w}}D_{\si\si}(\vec{k}-
\vec{\w})}{(\vec{k}-\vec{\w})^2}\frac{\imath}{2\pi}\lim_{R
\rightarrow\infty}\int_{-R}^{R}\frac{d\w_0}{\left[\w_0-{\cal C}_r+
\imath\e\right]}+{\cal O}\left(1/m\right)\nonumber\\
&=&m+\frac{1}{2}g^2C_F\int_r\frac{\dk{\vec{\w}}D_{\si\si}(\vec{k}-
\vec{\w})}{(\vec{k}-\vec{\w})^2}+{\cal O}\left(1/m\right).\label
{eq:const_ren}
\eea
The effect of performing the temporal integral first is that the 
regularized constant ${\cal C}_r$ in the denominator factor 
becomes irrelevant.  The spatial integral now involves no external 
scale and removing the spatial regularization, we arrive at our 
above result, \eq{eq:const}.

A brief discussion about the physical interpretation of these 
results is in order.  Firstly, it might be the case that there 
exist other solutions to the truncated gap equation.  As will be 
shown in the next Section though, the above solution can also be 
derived from a semi-perturbative type of expansion.  However, as 
is the case with many systems, especially those with strong 
couplings or phase transitions, the fully nonperturbative solution 
might not be the same as the resummed perturbative solution.  
Second, that the solution involves potentially divergent constants 
is not a comfortable situation but does not necessarily contradict 
the physics.  We notice that the quark propagator has a single 
pole, so cannot represent physical propagation (which requires a 
covariant double pole) and this arises obviously from the 
truncation of the mass expansion and where the charge conjugation 
symmetry has been explicitly broken.  Indeed, the position of the 
pole has no physical meaning since the quark can never be 
on-shell.  That the single pole is shifted to infinity simply 
means that either one requires infinite energy to create a quark 
from the vacuum or that should one have an incoming quark (from 
some other hadron), only the relative energy is important.  Note 
that in Coulomb gauge it is known that the total color charge of 
the system is conserved and vanishing \cite{Reinhardt:2008pr}, so 
one cannot prepare an isolated colored state of a single quark 
under any circumstances.  Further, the divergences here have no 
interpretation with regards to renormalization, at least within 
the context of the mass expansion to leading order.  The mass 
parameter cannot be renormalized simply because one cannot 
construct an appropriate counterterm in the action that is linear 
in $m$ to absorb the constant ${\cal C}$.  Also, the quark field 
renormalization is trivial at leading order, as one sees from the 
explicit form of the temporal quark-gluon vertex, \eq{eq:vsol}.  
The upshot of this is that one must consider only the relative 
energies in the system -- the divergence of the absolute energy 
has no physical meaning and this will become especially relevant 
when we study the \BS equation (indeed this has been known for 
quite some time, see for example Ref.~\cite{Adler:1984ri}).

Having discussed the quark propagator, let us now discuss the 
antiquark propagator.  Recall that at tree-level, we used a 
different Feynman prescription for the two denominator factors and 
this gives rise to some rather interesting physical consequences.  
As previously discussed, the heavy mass expansion employed here 
breaks the charge conjugation symmetry relating particle and 
antiparticle, so we cannot expect that the two propagators are 
necessarily equivalent.  Starting with the gap equation for full 
QCD, \eq{eq:gap}, we reverse the ordering of the quark and 
antiquark functional derivatives that form the quark Green's 
functions (still within the context of the full quark fields and 
sources) and rearrange the ordering to get the gap equation for 
the antiquark propagator:
\bea
-\G_{q\ov{q}\de\al}(-k)&=-\G_{q\ov{q}\de\al}^{(0)}(-k)-\int\dk{\w}&
\left\{\G_{q\ov{q}\si\de\ga}^{b}(-k,\w,k-\w)W_{q\ov{q}\ga\ba}(-\w)
\G_{q\ov{q}\si\ba\al}^{(0)a}(-\w,k,\w-k)W_{\si\si}^{ab}(k-\w)
\right.\nonumber\\&&\left.
+\G_{q\ov{q}A\de\ga j}^{b}(-k,\w,k-\w)W_{q\ov{q}\ga\ba}(-\w)\G_{q
\ov{q}A\ba\al i}^{(0)a}(-\w,k,\w-k)W_{AAij}^{ab}(k-\w)
\right\}.\nonumber\\
\label{eq:bgap}
\eea
Applying our truncation scheme reduces the above to
\be
\G_{q\ov{q}\de\al}(-k)=\G_{q\ov{q}\de\al}^{(0)}(-k)+\int\dk{\w}\G_{
q\ov{q}\si\de\ga}^{b}(-k,\w,k-\w)W_{q\ov{q}\ga\ba}(-\w)\G_{q\ov{q}
\si\ba\al}^{(0)a}(-\w,k,\w-k)W_{\si\si}^{ab}(k-\w)+{\cal O}
\left(1/m\right).
\label{eq:bglap1}
\ee
In similar fashion, we have the \ST identity for the 
antiquark-gluon vertex:
\be
-k_3^0\G_{q\ov{q}\si\ba\al}^{d}(k_2,k_1,k_3)=
+\G_{q\ov{q}\ba\de}(k_2)\left[igT^d\right]_{\de\al}^T-\left[igT^d
\right]_{\ba\de}^T\G_{q\ov{q}\de\al}(-k_1)+{\cal O}\left(1/m\right)
.
\ee
The form of the solution to \eq{eq:bglap1} is similar to the 
previous results:
\be
W_{q\ov{q}\al\ba}(k)=\frac{-\imath\de_{\al\ba}}{\left[k_0+
\ov{{\cal C}}_r+\imath\e\right]}+{\cal O}\left(1/m\right),\;\;\;\;
\G_{q\ov{q}\al\ba}(k)=\imath\de_{\al\ba}\left[k_0+\ov{{\cal C}}_r
\right]+{\cal O}\left(1/m\right)
\label{eq:bgapsol}
\ee
with the corresponding solution for the vertex
\be
\G_{q\ov{q}\si\al\ba}^{d}(k_1,k_2,k_3)=-\left[gT^d\right]_{\ba\al}+
{\cal O}\left(1/m\right).
\label{eq:bvsol}
\ee
However, the constant (obtained in the same way as before by first 
doing the temporal integral under the assumption that the spatial 
integral is regularized in some hypothetical manner) is now
\be
\ov{{\cal C}}_r=m-\frac{1}{2}g^2C_F\int_{r}\frac{\dk{\vec{\w}}D_{
\si\si}(\vec{\w})}{\vec{\w}^2}+{\cal O}\left(1/m\right).
\label{eq:bconst}
\ee
We notice that the sign of the loop correction has changed and 
this will turn out (in the context of the \BS equations for mesons 
and diquarks) to have far-reaching consequences.  For the moment 
we interpret this result as being simply another manifestation of 
the breaking of the charge conjugation (particle--antiparticle) 
symmetry via the mass expansion.

\section{Gap equation: semiperturbative treatment}
\setcounter{equation}{0}

\begin{figure}[t]
\vspace{0.5cm}
\includegraphics[width=0.33\linewidth]{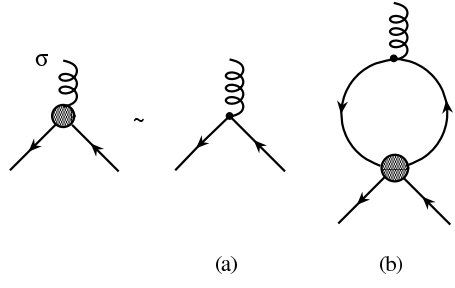}
\caption{\label{fig:qbqs}Diagrams that contribute under our 
truncation scheme to the \DS equation for the \emph{temporal} 
quark-gluon vertex (without prefactors or signs).  Internal 
propagators are fully dressed and blobs represent dressed proper 
vertex and (reducible) kernels.  Internal propagators represented 
by solid lines represent the quark propagator.}
\end{figure}

\begin{figure}[t]
\vspace{0.5cm}
\includegraphics[width=0.3\linewidth]{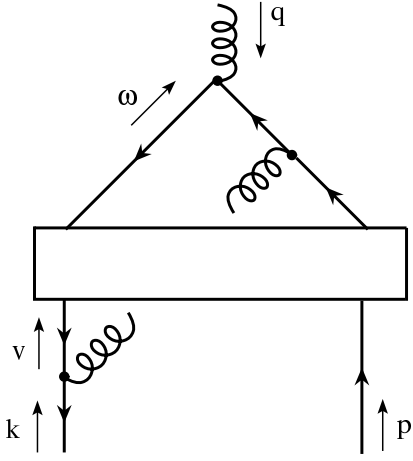}
\caption{\label{fig:qbqscbox}Generic crossed box type loop 
correction to the temporal quark-gluon vertex.  Internal 
propagators are fully dressed: solid lines represent the quark 
propagator and springs denote the temporal gluon propagator.  The 
box represents any combination of interactions allowed under our 
truncation.}
\end{figure}

Whilst we have already solved the gap (and anti-gap) equation, it 
proves instructive to reconsider it within the context of a 
semiperturbative analysis since this will naturally introduce a 
technical feature crucial for considering the \BS equation 
nonperturbatively.  We have seen that the solutions for the proper 
two-point function leads to a temporal quark-gluon (and 
antiquark-gluon) vertex that is not dressed.  Under our truncation 
scheme, the nonperturbative \DS equation for the temporal 
quark-gluon vertex involves the diagrams shown in 
Fig.~\ref{fig:qbqs}.  The semiperturbative expansion is based on a 
hybrid loop expansion whereby all internal propagators are taken 
to be dressed, but all internal vertices are tree-level.  To show 
that all the loop corrections (contained in diagram (b) of 
Fig.~\ref{fig:qbqs}) vanish, it suffices to consider two types of 
diagram, given in Figs.~\ref{fig:qbqslad} and \ref{fig:qbqscbox}.

In Fig.~\ref{fig:qbqslad} we consider a single ladder exchange
correction to the temporal quark-gluon vertex.  This diagram
(neglecting the overall color and prefactors) gives rise to the
following scalar integral (as in the previous section, the spatial
integral is regularized such that one can perform the temporal
integral first):
\begin{figure}[t]
\vspace{0.5cm}
\includegraphics[width=0.2\linewidth]{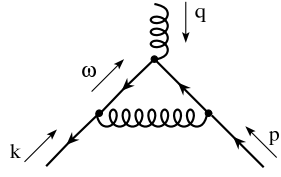}
\caption{\label{fig:qbqslad}Ladder type loop correction to the 
temporal quark-gluon vertex.  Internal propagators are fully 
dressed: solid lines represent the quark propagator and springs 
denote the temporal gluon propagator.}
\end{figure}
\be
\int_r\frac{\dk{\vec{\w}}D_{\si\si}(\vec{k}-\vec{\w})}{(\vec{k}-
\vec{\w})^2}\frac{1}{2\pi}\int_{-\infty}^{\infty}\frac{d\w_0}{\left
[\w_0-{\cal C}_r+\imath\e\right]\left[\w_0+q_0-{\cal C}_r+\imath\e
\right]}.
\ee
Now we apply the following identity (for finite, real $a,b$; the 
case $a=b$ is trivial):
\be
\int_{-\infty}^{\infty}\frac{dz}{\left[z-a+\imath\e\right]
\left[z-b+\imath\e\right]}=\frac{1}{(a-b)}\int_{-\infty}^{\infty}
dz\left\{\frac{1}{\left[z-a+\imath\e\right]}-\frac{1}{\left[z-b+
\imath\e\right]}\right\}=0.
\label{eq:intid}
\ee
Thus we see that where the spatial integral is regularized, the 
temporal integral vanishes.  It is simple to see that the planar 
one-loop diagrams with two or more external temporal gluon legs 
(which under the truncation scheme considered here connect only to 
the internal quark line) and one internal temporal gluon will also 
vanish.

Now let us consider a generic crossed box (nonplanar) type of 
diagram, illustrated in Fig.~\ref{fig:qbqscbox}.  Considering only 
the temporal double integral components of the explicit internal 
quark propagators, we have the following form
\bea
&&\int_{-\infty}^{\infty}\frac{d\w_0\,dv_0}{\left[v_0-a_1+\imath\e
\right]\left[\w_0-a_2+\imath\e\right]\left[\w_0+q_0-a_3+\imath\e
\right]\left[\w_0-v_0-p_0-a_4+\imath\e\right]}\nonumber\\&&
=\int_{-\infty}^{\infty}\frac{d\w_0}{\left[\w_0-a_2+\imath\e\right]
\left[\w_0+q_0-a_3+\imath\e\right]\left[\w_0-p_0-a_1-a_4+2\imath\e
\right]}\nonumber\\&&
\times\int_{-\infty}^{\infty}dv_0\left\{\frac{1}{\left[v_0-a_1+
\imath\e\right]}-\frac{1}{\left[v_0-\w_0+p_0+a_4-\imath\e\right]}
\right\}\nonumber\\&&
=-2\pi\imath\int_{-\infty}^{\infty}\frac{d\w_0}{\left[\w_0-a_2+
\imath\e\right]\left[\w_0+q_0-a_3+\imath\e\right]\left[\w_0-p_0-a_1
-a_4+2\imath\e\right]}\nonumber\\&&
=0
\eea
where in the last line, we have used a variation of the identity 
\eq{eq:intid}.  Thus we have the result that the generic crossed 
box type of diagram shown in Fig.~\ref{fig:qbqscbox} also vanishes.

Given that both the single ladder type exchange diagram and the 
generic crossed box diagrams considered so far vanish, it is easy 
to see that any vertex dressing diagram will vanish (including all 
subdiagrams such as internal vertex corrections and so on), since 
all diagrams are merely variations or combinations of these two 
under our truncation scheme.  This result is a consequence of the 
fact that the energy and Feynman prescription of the denominator 
factors follow the quark line through the diagram so that 
eventually the identity, \eq{eq:intid}, can be used.  It is also 
precisely the reason why all closed quark loops vanish, as 
previously discussed.

Thus, the semiperturbative expansion confirms our previous result 
that the temporal quark-gluon vertex remains bare to all orders.  
Also clear should be that the result applies to the 
antiquark-gluon vertex too.  With the corresponding simple forms 
for the self-energy integrals of the gap and anti-gap equations 
(which as we recall, reduce to the rainbow truncation), the 
results for the quark propagator functions are also confirmed.  
Notice though that whilst the all orders semiperturbative result 
must match the nonperturbative result, the converse is not 
necessarily true.  It remains the case that there may exist 
further solutions, but these must be purely nonperturbative in 
character if they exist.  That the identities that have been 
introduced here conform to the nonperturbative case previously 
studied is useful since it allows us to apply them with confidence 
to the \BS equation in the next Sections.

\section{\BS equation: mesons}
\begin{figure}[t]
\vspace{0.5cm}
\includegraphics[width=0.4\linewidth]{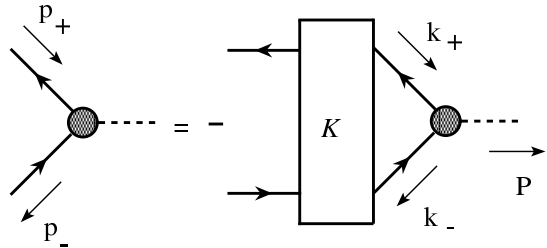}
\caption{\label{fig:fbse}Homogeneous \BS equation for 
quark-antiquark bound states.  Internal propagators are fully 
dressed and solid lines represent the quark propagator.  The box 
represents the \BS kernel $K$ and filled blobs represent the \BS 
vertex function $\G$ with the (external) bound state leg given by 
a dashed line.  See text for details.}
\end{figure}

\setcounter{equation}{0}
As has been emphasized, because we initially consider the 
generating functional for full quark sources and fields we are at 
liberty to use the full functional (nonperturbative) equations as 
a starting point and then subsequently apply our mass expansion 
and truncation scheme.  Let us now consider the homogeneous \BS 
equation for quark-antiquark bound states.  In full QCD and with 
our conventions (the minus sign arises from the definitions of the 
Legendre transform and Green's functions), this equation reads
\be
\G_{\al\ba}(p;P)=-\int\dk{k}K_{\al\ba;\de\ga}(p,k;P)\left[W_{\ov{q}
q}(k_+)\G(k;P)W_{\ov{q}q}(k_-)\right]_{\ga\de}
\ee
where $k_+=k+\xi P$, $k_-=k+(\xi-1)P$ (similarly for $p_\pm$) are 
the momenta of the quarks, $P$ is the pole 4-momentum of the bound 
state (assuming that a solution exists), $\xi=[0,1]$ is the 
so-called momentum sharing fraction that dictates how much of the 
total meson momentum is carried by each quark constituent, $K$ 
represents the \BS kernel and $\G$ is the \BS vertex function for 
the particular bound state that one is considering and whose 
indices explicitly denote only its quark content 
(we will see later what color, flavor and spin structure the 
solutions may have). 
Physically, results should be independent of $\xi$ and this has 
been numerically observed in phenomenological studies 
\cite{Alkofer:2002bp}.  The \BS equation is shown pictorially in 
Fig.~\ref{fig:fbse}.

Aside from the quark propagators (which we shall discuss shortly), 
the central element to solving the \BS equation involves the 
construction of the kernel $K$.  As discussed in the Introduction, 
for technical reasons the most widely studied system is based on 
the ladder kernel which is either constructed via the interchange 
of a single gluon (for example \cite{Maris:1999nt}) or as a 
phenomenological potential (see for example 
Ref.~\cite{Adler:1984ri}).  However, there has been much
recent attention focused on the construction of more sophisticated
kernels.  One key element of the construction is the axialvector
Ward--Takahashi identity [AXWTI], which relates the gap equation to
the \BS kernel and which ensures that chiral symmetry and its
spontaneous breaking are consistently implemented (e.g., 
Refs.~\cite{Adler:1984ri,Bender:1996bb,Watson:2004kd}).  Here, we 
shall show that the ladder \BS kernel is \emph{exact} at leading 
order in the heavy mass expansion and under our truncation scheme.
This derivation follows in the same way as the semiperturbative 
analysis of the previous Section.

Following from the truncation of the heavy mass expansion to 
leading order, it was argued that the antiquark propagator must be 
treated as distinct from the quark propagator.  This means that we 
have to be very explicit about which propagator is which.  Since 
we are studying the quark-antiquark system (we will analyze the 
quark-quark, or diquark system in the next Section), the \BS 
equation for our purposes more properly reads
\be
\G_{\al\ba}(p;P)=-\int\dk{k}K_{\al\ba;\de\ga}(p,k;P)\left[W_{\ov{q}
q}(k_+)\G(k;P)(-1)W_{q\ov{q}}^T(-k_-)\right]_{\ga\de}
\ee
where we have explicitly identified the antiquark propagator 
contribution (it corresponds to the lower line of 
Fig.~\ref{fig:fbse}) by reordering the functional derivatives and 
there is implicitly a similar antiquark contribution within the 
kernel which absorbs the explicit minus sign.  Recall that we are 
implicitly considering only the flavor non-singlet case.

\begin{figure}[t]
\vspace{0.5cm}
\includegraphics[width=0.3\linewidth]{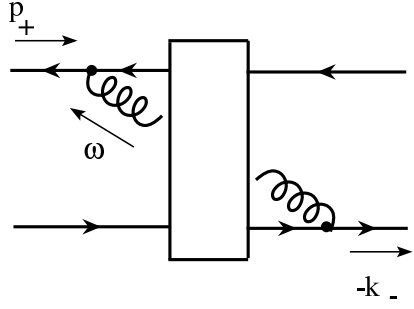}
\caption{\label{fig:bscbox}Generic crossed box type of diagram 
that contributes to the \BS kernel. Internal propagators are fully 
dressed, whereas vertices are tree-level.  The upper (lower) solid 
line denotes the quark (antiquark) propagator; springs denote the 
temporal gluon propagator and the box represents any combination 
of nontrivial interactions allowed under our truncation scheme.  
See text for details.}
\end{figure}

To construct the full \BS kernel, consider the generic 
semiperturbative crossed box contribution given in 
Fig.~\ref{fig:bscbox}.
As before, at leading order in the mass expansion and under our 
truncation scheme we only have the temporal quark-gluon and 
antiquark-gluon vertices, both of which have been shown to be 
given by their tree-level forms.  Additionally, it was justified 
that all planar diagrams with multiple external temporal gluon 
legs vanish.  Thus, aside from the ladder contribution to the \BS 
kernel, this generic crossed box diagram contains all possible 
nontrivial contributions.  Such a diagram has at least the 
following terms in the temporal integral (as before, we assume 
that the spatial integral is regularized and finite so that we are 
able to firstly perform the temporal integral without complication)
\be
\int\frac{d\w_0}{\left[\w_0+p_{+}^0-{\cal C}_r+\imath\e\right]
\ldots\left[\w_0-k_{-}^0+\ov{{\cal C}}_r+\imath\e\right]}.
\ee
The first factor corresponds to the explicit quark (upper) 
propagator, the last factor to the explicit antiquark (lower) 
propagator.  Implicitly (represented by the dots), there may be 
multiple propagator factors which carry the same dependence on the 
integration energy $\w_0$ but crucially, all these will have the 
same relative sign for the Feynman prescription term, i.e., 
$\w_0+\imath\e$, regardless of whether they originate from 
internal quark or antiquark propagators.  Therefore, this type of 
integral can always be reduced to the difference of integrals over 
a simple pole and with the same sign for carrying out the analytic 
integration, just as in \eq{eq:intid} and the semiperturbative 
treatment of the vertex discussed in the last Section.  Thus, all 
generic crossed box diagrams in the \BS kernel are zero and one is 
left with simply the ladder contribution to the kernel.  Actually, 
this could have been anticipated from the beginning -- the AXWTI 
connects the self-energy term of the gap equation and the \BS 
kernel and since it has been shown explicitly that the self-energy 
integral reduces to the rainbow truncation, the corresponding \BS 
kernel is simply given by ladder exchange.

Given all this, the \BS equation for the quark-antiquark system, 
at leading order in the mass expansion and within our truncation 
scheme, can be explicitly written as
\bea
\G_{\al\ba}(p;P)&=&-\int\dk{k}\G_{\ov{q}q\si\al\ga}^{a}(p_+,-k_+,
k-p)W_{\si\si}^{ab}(p-k)\G_{q\ov{q}\si\ba\de}^{bT}(-p_-,k_-,p-k)
W_{\ov{q}q\ga\la}(k_+)W_{q\ov{q}\de\ka}^T(-k_-)\G_{\la\ka}(k;P)
\nonumber\\
&&+{\cal O}\left(1/m\right).
\eea
Inserting the nonperturbative results for the propagators and 
vertices
so far,
Eqs.~(\ref{eq:gapsol},\ref{eq:bgapsol},\ref{eq:vsol},
\ref{eq:bvsol})
and taking the form, \eq{eq:gtemp}, for the temporal gluon 
propagator,
we get the equation (the temporal and spatial integrals are 
separated as before)
\be
\G_{\al\ba}(p;P)=g^2\int_r\frac{\dk{\vec{k}}D_{\si\si}(\vec{p}-\vec
{k})}{(\vec{p}-\vec{k})^2}\frac{\imath}{2\pi}\int_{-\infty}^\infty
\frac{dk_0}{\left[k_+^0-{\cal C}_r+\imath\e\right]\left[k_-^0-\ov{{
\cal C}}_r-\imath\e\right]}\left[T^a\G(k;P)T^a\right]_{\al\ba}+
{\cal O}\left(1/m\right).
\ee
We see immediately that the flavor and spin structure of the meson 
decouples from the problem -- this is well a known property of the 
heavy mass expansion.  The color structure will be discussed 
shortly.  Since the external energy $p_0$ does not enter the 
right-hand side of the above equation, we further have that the 
\BS equation must be independent of the relative quark energy (and 
only implicitly dependent on the putative bound state energy $P_0$)
.  Thus we can write
\bea
\G_{\al\ba}(\vec{p};P)&=&g^2\int_r\frac{\dk{\vec{k}}D_{\si\si}(\vec
{p}-\vec{k})}{(\vec{p}-\vec{k})^2}\left[T^a\G(\vec{k};P)T^a\right]_
{\al\ba}\frac{\imath}{2\pi}\int_{-\infty}^\infty\frac{dk_0}{\left[
k_+^0-{\cal C}_r+\imath\e\right]\left[k_-^0-\ov{{\cal C}}_r-\imath
\e\right]}+{\cal O}\left(1/m\right)\nonumber\\
&=&-g^2\int_r\frac{\dk{\vec{k}}D_{\si\si}(\vec{p}-\vec{k})}{(\vec
{p}-\vec{k})^2}\frac{\left[T^a\G(\vec{k};P)T^a\right]_{\al\ba}}{
\left[P_0-{\cal C}_r+\ov{{\cal C}}_r+2\imath\e\right]}+{\cal O}
\left(1/m\right).
\label{eq:fpaux}
\eea
Thus, at leading order in the mass expansion, inserting the
expressions \eq{eq:const_ren} and \eq{eq:bconst}, for ${\cal C}_r$ 
and
$\ov{{\cal C}}_r$, respectively, it is now clear that
\be
\left[P_0-g^2C_F\int_{r}\frac{\dk{\vec{\w}}D_{\si\si}(\vec{\w})}
{\vec{\w}^2}\right]\G_{\al\ba}(\vec{p};P)=-g^2\int_{r}\frac{\dk{
\vec{k}}D_{\si\si}(\vec{p}-\vec{k})}{(\vec{p}-\vec{k})^2}\left[T^a
\G(\vec{k};P)T^a\right]_{\al\ba}+{\cal O}\left(1/m\right).
\ee
We notice that the explicit quark mass contributions of the 
self-energy expressions cancel.  This is a feature of the 
quark-antiquark \BS equation -- it does not make any reference to 
the origins of its constituents and why for example the pion can 
be a massless bound state of massive constituents.  Physically, 
one can visualize that the quark and antiquark are moving with 
equal and opposite velocities such that the centre of mass system 
(the bound state) is stationary.  This is related explicitly to 
the choice of Feynman prescription for the constituent quark and 
antiquark.  Were the Feynman prescription for the antiquark chosen 
to coincide with that of the quark, the right-hand side of 
\eq{eq:fpaux} would simply vanish and there would be certainly no 
physical quark-antiquark state.  The Feynman prescription for the 
antiquark corresponds precisely to a particle moving with the 
opposite velocity.  Also, at leading order in the mass expansion, 
the momentum sharing parameter, $\xi$, has dropped out.  Thus, the 
results retain the important physical requirement that they be 
independent of $\xi$.  Shifting integration momenta, we can write
\be
P_0\G_{\al\ba}(\vec{p};P)=g^2\int_{r}\frac{\dk{\vec{\w}}D_{\si\si}
(\vec{\w})}{\vec{\w}^2}\left\{C_F\G_{\al\ba}(\vec{p};P)-\left[T^a\G
(\vec{p}-\vec{\w};P)T^a\right]_{\al\ba}\right\}+{\cal O}\left(1/m
\right).
\ee
To see the physical meaning of this equation, we rewrite the \BS 
vertex function as a Fourier transform:
\be
\G_{\al\ba}(\vec{p};P)=\int d\vec{y}e^{-\imath\vec{p}\cdot\vec{y}}
\G_{\al\ba}(\vec{y})
\ee
(in the homogeneous \BS equation, the total momentum $P$ denotes 
the solution and is not a variable).  We also write for the color 
structure
\be
\left[T^a\G(\vec{y})T^a\right]_{\al\ba}=C_M\G_{\al\ba}(\vec{y})
\ee
where $C_M$ is yet to be identified.  Then, the \BS equation 
reduces to
\be
\int d\vec{y}e^{-\imath\vec{p}\cdot\vec{y}}P_0\G_{\al\ba}(\vec{y})=
\int d\vec{y}e^{-\imath\vec{p}\cdot\vec{y}}g^2\int_{r}\frac{\dk{
\vec{\w}}D_{\si\si}(\vec{\w})}{\vec{\w}^2}\left\{C_F\G_{\al\ba}(
\vec{y})-e^{\imath\vec{\w}\cdot\vec{y}}C_M\G_{\al\ba}(\vec{y})
\right\}+{\cal O}\left(1/m\right)
\ee
with the simple solution
\be
P_0=g^2\int_{r}\frac{\dk{\vec{\w}}D_{\si\si}(\vec{\w})}{\vec{\w}^2}
\left\{C_F-e^{\imath\vec{\w}\cdot\vec{y}}C_M\right\}+{\cal O}
\left(1/m\right).
\ee
Because the total color charge of the system is conserved and
vanishing \cite{Reinhardt:2008pr}, neither the quark or antiquark 
can exist as an independent asymptotic physical state.  Thus, the 
bound state energy, $P_0$, can only increase linearly as the 
separation between the them increases (physically confining) or be 
infinite when the hypothetical regularization is removed (so that 
the system  cannot be physically created).  Whether the system is 
confining or disallowed can only depend on the color structure, 
since the temporal gluon propagator dressing function would be 
common to both situations.  In configuration space, an infrared 
confining solution is characterized by the solution 
$P_0=\si|\vec{y}|$ for large $|\vec{y}|$ and where $\si$ is the 
string tension such that as the separation between the quark and 
antiquark increases, the energy of the system should increase 
linearly without bound and infinite energy input is required to 
fully separate them (at least in the absence of unquenching).  The 
small $|\vec{y}|$ (and large $|\vec{\w}|$) properties are of no 
concern here.  The Fourier transform integral needed for the 
infrared confining solution is
\be
\int\frac{\dk{\vec{\w}}}{\vec{\w}^4}\left[1-e^{\imath\vec{\w}\cdot
\vec{y}}\right]=\frac{|\vec{y}|}{8\pi}.
\ee
If the temporal gluon propagator dressing function is more 
infrared divergent than $1/|\vec{\w}|$, as above, then 
\be
C_F=C_M
\ee
is required such that the spatial integral is convergent and the
energy of the system well-defined, as the hypothetical 
regularization is removed (since we are interested in the low 
$|\vec\w|$ regime, it becomes clear that the  regularization here 
would be infrared in character).  Using the Fierz identity for the 
generators $T^a$:
\be
2\left[T^a\right]_{\al\ba}\left[T^a\right]_{\de\ga}=\de_{\al\ga}
\de_{\de\ba}-\frac{1}{N_c}\de_{\al\ba}\de_{\de\ga}
\label{eq:fierz}
\ee
gives the condition
\be
C_F\G_{\al\ga}(\vec{y})\equiv C_M\G_{\al\ga}(\vec{y})=\frac{1}{2}
\de_{\al\ga}\G_{\ba\ba}(\vec{y})-\frac{1}{2N_c}\G_{\al\ga}(\vec{y})
,\ee
or with the definition \eq{eq:casimir1},
\be
\G_{\al\ga}(\vec{y})=\de_{\al\ga}\G(\vec{y}).
\ee
In other words, the quark-antiquark \BS equation with $D_{\si\si}$ 
more infrared singular than $1/|\vec{\w}|$ can only have a finite 
solution for \emph{color singlet} states where the divergent 
constant integral coming from the unphysical quark self-energy 
cancels; otherwise the energy of the system is divergent.

Another way to see that only color-singlet states are physical is 
to consider that if the temporal gluon propagator is the origin of 
a potential in configuration space, this potential can always be 
shifted by some spatial constant \cite{Adler:1984ri}.  In momentum 
space, this means that one can make the replacement
\be
g^2\frac{D_{\si\si}(\vec{\w})}{\vec{\w}^2}\rightarrow g^2\frac{D_{
\si\si}(\vec{\w})}{\vec{\w}^2}+\mbox{const}\times(2\pi)^3\de(\vec{
\w})
\ee
without changing the physical bound state energy.  This 
automatically gives the constraint $C_F=C_M$ regardless of whether 
or not the temporal gluon propagator is infrared enhanced.

Assuming that in the infrared (as is indicated by the lattice data 
\cite{Quandt:2008zj} or by the above argument about the 
non-existence of asymptotic quark states), 
$D_{\si\si}=X/\vec{\w}^2$ where $X$ is some combination of 
constants (and further knowing that $g^2X$ is a renormalization 
group invariant \cite{Zwanziger:1998ez,Watson:2008fb}), then
\be
P_0\equiv\si|\vec{y}|=\frac{g^2C_FX}{8\pi}|\vec{y}|+{\cal O}
\left(1/m\right).
\ee
The above result is that there exists a direct connection between 
the string tension and the nonperturbative Yang--Mills sector of 
QCD at least under the truncation scheme considered here.  The 
veracity of the truncation scheme will be discussed at the end.

\section{\BS equation: diquarks}
\setcounter{equation}{0}
Let us now briefly take a look at the diquark \BS equation.  The 
difference between this and the previously considered 
quark-antiquark system is actually rather simple from a technical 
standpoint, but leads to a completely different physical result.  
Recall that the quark and the antiquark propagators share the same 
Feynman prescription relative to their energy and only the 
constant components change.  Thus, preserving the momentum routing 
of the \BS equation, the result that the crossed box contributions 
to the \BS kernel extends to the diquark case since this is purely 
dependent on the Feynman prescription.  This means that we can 
immediately write down the \BS equation for diquarks, at leading 
order in the mass expansion and within our truncation scheme:
\bea
\G_{\al\ba}(p;P)&=&-\int\dk{k}\G_{\ov{q}q\si\al\ga}^{a}(p_+,-k_+,
k-p)W_{\si\si}^{ab}(p-k)\G_{\ov{q}q\si\ba\de}^{b}(-p_-,k_-,p-k)W_{
\ov{q}q\ga\la}(k_+)W_{\ov{q}q\de\ka}(-k_-)\G_{\la\ka}(k;P)
\nonumber\\&&
+{\cal O}\left(1/m\right).
\eea
Again, the indices of the \BS vertex function correspond to the 
quark content of the diquark and since the flavor and spin content 
decouple from the system, we shall only be interested in the color 
content of the diquark below.  Expanding this out as before, we 
get the analogous result
\be
\left[P_0-2m-g^2C_F\int_{r}\frac{\dk{\vec{\w}}D_{\si\si}(\vec{\w})}
{\vec{\w}^2}\right]\G_{\al\ba}(\vec{p};P)=g^2\int_{r}\frac{\dk{\vec
{\w}}D_{\si\si}(\vec{\w})}{\vec{\w}^2}\left[T^a\right]_{\al\la}
\left[T^a\right]_{\ba\ka}\G_{\la\ka}(\vec{p}-\vec{\w};P)+{\cal O}
\left(1/m\right).
\ee
Fourier transforming as previously, and writing
\be
\left[T^a\right]_{\al\la}\left[T^a\right]_{\ba\ka}\G_{\la\ka}(\vec{
y})=C_D\G_{\al\ba}(\vec{y})
\ee
gives the solution
\be
P_0=2m+g^2\int_{r}\frac{\dk{\vec{\w}}D_{\si\si}(\vec{\w})}{\vec{\w}
^2}\left\{C_F+e^{\imath\vec{\w}\cdot\vec{y}}C_D\right\}+{\cal O}
\left(1/m\right).
\ee
The reappearance of the quark mass simply indicates that in 
contrast to the quark-antiquark system, there are now two 
co-moving quarks.  The expression for the anti-diquark system is 
identical to the above, but with \emph{minus} twice the mass -- 
their velocities are simply reversed.  The diquark is 
antisymmetric under interchange of the two quark legs and since 
the spin and flavor structure decouples, this means that the color 
structure must be antisymmetric.  As for the quark-antiquark 
system, the system can only have a finite energy confining (energy 
increasing with separation) solution (if $C_D=-C_F$) or no finite 
solution at all.  Using the Fierz identity \eq{eq:fierz}, the 
color condition then reads
\be
-C_F\G_{\al\ba}(\vec{y})\equiv C_D\G_{\al\ba}(\vec{y})=\frac{1}{2}
\G_{\ba\al}(\vec{y})-\frac{1}{2N_c}\G_{\al\ba}(\vec{y}).
\ee
Demanding the diquark color antisymmetry and with the definition 
\eq{eq:casimir1} this becomes
\be
N_c^2-N_c-2=0\;\;\;\;\Rightarrow\;\;\;\;N_c=-1,2.
\ee
So in $SU(N_c=2)$ there exists a confined, antisymmetric bound 
state of two quarks -- the $SU(2)$ baryon and otherwise there are 
no (finite) physical states.

\section{Summary, discussion and conclusions}
\setcounter{equation}{0}

The connection between the Green's functions of Coulomb gauge \YM 
theory and physical quark confinement has been studied.  Since all 
quarks are confined (irrespective of their mass), the generating 
functional of full QCD was expanded in the mass parameter and the 
leading order considered in a manner appropriate to Coulomb gauge 
and such that the system simplifies dramatically.  This allowed 
the usage of the full nonperturbative quark equations of QCD in 
order to study the confinement properties.  Using the Coulomb 
gauge \ST identity for the quark-gluon vertices (whose derivation 
is one of the results of this study) and truncating the \YM sector 
to include only the nonperturbative gluon propagator, it was shown 
how the rainbow approximation to the quark and antiquark gap 
equations is exact in this case.  It was then demonstrated that 
the corresponding ladder approximation to the \BS equation was 
also exact.  The mass expansion breaks the time-reversal and 
charge conjugation properties of the theory and in order to 
describe physical quark-antiquark states, the relevant Feynman 
prescription was introduced.  With the analytic solutions to the 
gap equation, the \BS equation was solved for the quark-antiquark 
and quark-quark channels.  It was found that the only solutions 
correspond to confinement, namely that only color-singlet meson 
and $SU(2)$ baryon states have finite energy.  This energy must 
increase linearly with separation either because: (i) there are no 
asymptotic quark states as is known in Coulomb gauge from the 
conserved and vanishing total charge \cite{Reinhardt:2008pr}, or 
(ii) as is suggested from the lattice results 
\cite{Quandt:2008zj}, the temporal propagator is infrared 
enhanced.  Further, there exists a direct connection between the 
temporal gluon propagator of \YM theory and the string tension, at 
least within this truncation scheme.

Because of the (almost embarrassingly) simple results, it is worth 
re-emphasizing the input to the calculations.  Different areas of 
study have been combined here: the heavy quark mass expansion, the 
\DS formalism along with the \ST identity and the \BS equation, 
all of which have been considered within Coulomb gauge.  Indeed, 
the simplification arising from this combination underscores how 
powerful the choice of Coulomb gauge is for nonperturbative 
studies.  There are two approximations used.  Firstly, the 
temporal gluon propagator is considered to be energy-independent.  
This is consistent with available lattice data 
\cite{Quandt:2008zj}, although explicitly in disagreement with 
perturbation theory \cite{Watson:2007vc}.  Since this study is 
interested in the infrared properties of the theory, using the 
lattice suggestion seems justified.  Actually, the connection 
between the explicit one-loop perturbative expressions and the 
leading order perturbative expansion (in covariant gauges, the 
resummation of the leading logarithms via the anomalous 
dimensions) is not yet understood in Coulomb gauge.  Further, 
since the nonperturbative ghost propagator in Coulomb gauge is 
strictly independent of the energy \cite{Watson:2008fb} in order 
to cancel the closed ghost-loops occurring in the \YM expressions, 
the temporal gluon propagator must have some part that is also 
independent of the energy (the spatial gluon propagator is 
explicitly dependent on the energy).  Thus it seems reasonable to 
assume that whatever energy dependence the temporal gluon 
propagator does have (and that is not visible with current lattice 
configurations), it is not important to the conclusions here.

The second approximation used in this study was to neglect the 
vertex and higher $n$-point functions of \YM theory.  Recall that 
the tree-level spatial quark-gluon vertex is suppressed by the 
mass.  This means that the number of truncated terms at leading 
order in the mass expansion is heavily suppressed.  It was seen 
that without such terms, the quark-gluon vertices (and the \BS 
kernel) reduced to their tree-level forms.  This suggests that 
whilst the dressing function $D_{\si\si}$ does implicitly contain 
all nonperturbative effects associated with the dynamical dressing 
of the color charge (including, for example, potential glueball 
states), the quark-gluon vertices correspond to the naked quark 
color charge.  One might express the results here as corresponding 
to a dressed color string confining two naked color sources.  In 
the gap and \BS equations, the explicit \YM contributions 
presumably cannot obliterate the effect of the rainbow-ladder 
exchange for a given temporal propagator input since (assuming the 
infrared enhancement as suggested from the lattice
\cite{Quandt:2008zj}) one would require an exact `anti-confining'
cancellation.  Covariant gauge studies of the effect of non-Abelian
corrections to the \BS equation as applied to the light quark 
sector indicate that they are dominant compared to the Abelian 
corrections \cite{Williams:2009wx}.  However, in the case studied 
here, there are no Abelian corrections and the diagrammatic 
content of the non-Abelian contribution is reduced by the mass 
expansion, making a quantitative comparison to known \BS studies 
somewhat speculative.  Perhaps a better comparison about the 
nature of neglecting the non-Abelian corrections is with 
Ref.~\cite{Zwanziger:1998ez}, which refers to the Wilson loop of 
pure \YM theory.  With effectively the same truncation (i.e., 
neglecting the \YM vertices), it was demonstrated that a string
tension (the so-called Coulomb string tension) can also be 
extracted. Subsequently, it was shown that the physical string 
tension is lower than the Coulomb string tension 
\cite{Zwanziger:2002sh} (a statement colloquially referred to as 
 ``no confinement without Coulomb confinement'').  Thus, we 
anticipate that the effect of including the non-Abelian 
corrections to the formalism presented here would result not in 
the removal of the linearly rising bound state energy (i.e., not 
the cancellation of the ladder exchange) but rather in the 
softening of the coefficient by shifting the pole position by some 
finite amount.  Physically, this amounts to the charge screening 
of the quark color charge.

The outlook of this study is extremely positive, with several 
avenues that might prove fruitful for further investigation.  
Maintaining the truncation scheme, one can consider the three 
quark (or two quark plus antiquark etc.) Faddeev equation to 
investigate $SU(3)$ baryons and whether or not the simple results 
for color confinement persist to the many-body system.  Even more 
speculatively, given the simplicity of the leading order mass 
expansion one might imagine going even further and considering the 
two quark, two antiquark system.  A second important line of study 
would be to consider the explicit inclusion of the \YM vertices to 
the combined system of Dyson--Schwinger, \ST and \BS equations, 
either phenomenologically or self-consistently to assess the 
influence of the truncation scheme considered so far.  A third 
possibility is to investigate what happens when one chooses a 
different velocity parameter in the original decomposition of the 
quark fields.  This would mix the temporal and spatial parts of 
the Coulomb gauge framework and allow one to investigate the 
connection between the respective gluon propagators.

\begin{acknowledgments}
The authors would like to thank G.~Burgio for useful discussions. 
CP has been supported by the Deutscher Akademischer Austausch 
Dienst (DAAD). PW and HR have been supported by the Deutsche 
Forschungsgemeinschaft (DFG) under contracts no. DFG-Re856/6-2,3. 
\end{acknowledgments}

\appendix

\section{Quark Slavnov-Taylor identity}
\label{app:stid}
\setcounter{equation}{0}

The derivation of the \ST identity for the quark-gluon vertex will 
be presented in this Appendix.  This derivation follows almost 
automatically as an extension to the \ST identities for the \YM 
sector \cite{Watson:2008fb}.  The full QCD action in the standard, 
second order formalism reads
\bea
{\cal S}_{QCD}=\int d^4x\left\{
\ov{q}_\al\left[\imath\ga^0\partial_{0}+gT^a\ga^0\si^a+\imath\s{
\vec{\ga}}{\div}-gT^a\s{\vec{\ga}}{\vec{A}^a}-m\right]_{\al\ba}q_
\ba-\frac{1}{4}F_{\mu\nu}^aF^{a\mu\nu}
\right\}
\eea
where the (antisymmetric) field strength tensor $F$ is given in 
terms of the gauge field $A_{\mu}^a$:
\be
F_{\mu\nu}^a=\partial_{\mu}A_{\nu}^a-\partial_{\nu}A_{\mu}^a+gf^
{abc}A_{\mu}^bA_{\nu}^c.
\ee
The action is invariant under a local $SU(N_c)$ gauge transform 
characterized by the parameter $\th_x^a$:
\be
U_x=\exp{\left\{-\imath\th_x^aT^a\right\}}
\ee
such that for infinitesimal $\th_x^a$, the fields transform as 
(recall that $\si\equiv A_0$)
\bea
\si_a&\rightarrow&\si'^a=\si^a-\frac{1}{g}\partial_0\th^a-f^{abc}
\si^b\th^c,\nonumber\\
\vec{A}^a&\rightarrow&\vec{A}'^a=\vec{A}^a+\frac{1}{g}\div\th^a-f^
{abc}\vec{A}^b\th^c,\nonumber\\
q_\al&\rightarrow&q'_\al=q_\al-\imath\th^a\left[T^a\right]_{\al\ba}
q_\ba,\nonumber\\
\ov{q}_\al&\rightarrow&\ov{q}'_\al=\ov{q}_\al+\imath\th^a\ov{q}_\ba
\left[T^a\right]_{\ba\al}.
\eea
Fixing to Coulomb gauge ($\s{\div}{\vec{A}^a}=0$) via the 
Faddeev-Popov technique introduces new terms into the action:
\be
{\cal S}_{FP}=\int d^4x\left[-\la^a\s{\div}{\vec{A}^a}-\ov{c}^a\s{
\div}{\vec{D}^{ab}}c^b\right],
\ee
where
\be
\vec{D}^{ab}=\de^{ab}\div-gf^{acb}\vec{A}^c
\ee
is the spatial covariant derivative (in the adjoint 
representation), $\la^a$ is a Lagrange multplier field to locally 
implement the gauge condition, $\ov{c}^a$ and $c^b$ are the 
Grassmann-valued ghost fields.  The action is invariant under a 
Gauss-BRST transform \cite{Zwanziger:1998ez} whereby the 
infinitesimal spacetime-dependent parameter $\th_x^a$ is 
factorized into two Grassmann-valued components: 
$\th_x^a=c_x^a\de\la_t$, where $\de\la_t$ is the 
\emph{time-dependent} infinitesimal variation.  The Gauss-BRST 
transform is a Coulomb gauge specific form of the standard BRS 
transform, allowed because the gauge-fixing term does not involve 
temporal differential operators.  The variations of the new fields 
read:
\be
\de\ov{c}^a=\frac{1}{g}\la^a\de\la_t,\;\;\;\;\de c^a=-\frac{1}{2}
f^{abc}c^bc^c\de\la_t,\;\;\;\;\de\la^a=0.
\ee
Including a source term,
\be
{\cal S}_s=\int d^4x\left[\ro^a\si^a+\s{\vec{J}^a}{\vec{A}^a}+
\ov{c}^a\et^a+\ov{\et}^ac^a+\xi^a\la^a+\ov{q}_\al\chi_\al+\ov{\chi}
_\al q_\al\right],
\ee
the generating functional is given by
\be
Z[J]=\int{\cal D}\Phi\exp{\left\{\imath{\cal S}_{QCD}+
\imath{\cal S}_{FP}+\imath{\cal S}_s\right\}}.
\ee
The Coulomb gauge \ST identities arise from regarding the 
Gauss-BRST transform as a change of integration variables under 
which the generating functional is invariant.  Since the Jacobian 
factor is trivial \cite{Watson:2006yq} and only the source term 
varies, we deduce that
\bea
0&=&\left.\int{\cal D}\Phi\frac{\de}{\de\left[\imath\de\la_t\right]
}\exp{\left\{\imath{\cal S}_{QCD}+\imath{\cal S}_{FP}+
\imath{\cal S}_s+\imath\de{\cal S}_s\right\}}\right|_{\de\la_t=0}
\nonumber\\
&=&\int{\cal D}\Phi\exp{\left\{\imath{\cal S}_{QCD}+\imath{\cal S}
_{FP}+\imath{\cal S}_s\right\}}\int d^4x\de(t-x_0)\left\{
-\frac{1}{g}\left(\partial_x^0\ro_x^a\right)c_x^a+f^{abc}\ro_x^a
\si_x^bc_x^c
\right.\nonumber\\&&\left.
-\frac{1}{g}J_{ix}^a\nabla_{ix}c_x^a+f^{abc}J_{ix}^aA_{ix}^bc_x^c-
\imath\ov{\chi}_{\al x}c_x^aT_{\al\ba}^aq_{\ba x}-\imath c_x^a
\ov{q}_{\ba x}T_{\ba\al}^a\chi_{\al x}+\frac{1}{g}\la_x^a\et_x^a+
\frac{1}{2}f^{abc}\ov{\et}_x^ac_x^bc_x^c
\right\}.
\eea
Just as for the pure \YM case, the time dependence of the 
variation, $\de\la_t$, results in the $\de$-function constraint 
$\de(t-x_0)$.  The above identity is most usefully expressed in 
terms of proper functions and repeating the manipulations of 
Ref.~\cite{Watson:2008fb}, one arrives at the identity
\bea
0&=\int d^4x\de(t-x_0)&\left\{
\frac{1}{g}\left(\partial_x^0\ev{\imath\si_x^a}\right)c_x^a
-f^{abc}\ev{\imath\si_x^a}\left[\ev{\imath\ro_x^b\imath\ov{\et}_
x^c}+\si_x^bc_x^c\right]
-\frac{1}{g}\left[\frac{\nabla_{ix}}{(-\nabla_x^2)}\ev{\imath A_
{ix}^a}\right]\ev{\imath\ov{c}_x^a}
\right.\nonumber\\&&
-f^{abc}\ev{\imath A_{ix}^a}t_{ij}(\vec{x})\left[\ev{\imath J_{jx}
^b\imath\ov{\et}_x^c}+A_{jx}^bc_x^c\right]
-\frac{1}{g}\la_x^a\ev{\ov{c}_x^a}
+\frac{1}{2}f^{abc}\ev{\imath c_x^a}\left[\ev{\imath\ov{\et}_x^b
\imath\ov{\et}_x^c}+c_x^bc_x^c\right]
\nonumber\\&&\left.
+\imath T_{\al\ba}^a\ev{\imath q_{\al x}}\left[\ev{\imath\ov{\chi}
_{\ba x}\imath\ov{\et}_x^a}-c_x^aq_{\ba x}\right]
+\imath T_{\ba\al}^a\left[\ev{\imath\chi_{\ba x}\imath\ov{\et}_x^a}
+c_x^a\ov{q}_{\ba x}\right]\ev{\imath\ov{q}_{\al x}}
\right\}.
\eea
Note that functional derivatives involving the Lagrange multiplier
result merely in a trivial identity such that the classical field 
$\la_x^a$ can be set to zero \cite{Watson:2008fb}.  Further, to 
derive the quark \ST identities, one functional derivative with 
respect to $\imath c_z^d$ is needed and then the ghost 
fields/sources can be set to zero.  Implementing this then gives
\bea
0&=\int d^4x\de(t-x_0)&\left\{
-\frac{\imath}{g}\left(\partial_x^0\ev{\imath\si_x^d}\right)
\de(z-x)
-f^{abc}\ev{\imath\si_x^a}\left[\frac{\de}{\de\imath c_z^d}
\ev{\imath\ro_x^b\imath\ov{\et}_x^c}-\imath\si_x^b\de^{dc}\de(z-x)
\right]
\right.\nonumber\\&&
+\frac{1}{g}\left[\frac{\nabla_{ix}}{(-\nabla_x^2)}\ev{\imath 
A_{ix}^a}\right]\ev{\imath\ov{c}_x^a\imath c_z^d}
-f^{abc}\ev{\imath A_{ix}^a}t_{ij}(\vec{x})\left[\frac{\de}{\de
\imath c_z^d}\ev{\imath J_{jx}^b\imath\ov{\et}_x^c}-\imath A_{jx}^b
\de^{dc}\de(z-x)\right]
\nonumber\\&&
-\imath T_{\al\ba}^a\ev{\imath q_{\al x}}\left[\frac{\de}{\de
\imath c_z^d}\ev{\imath\ov{\chi}_{\ba x}\imath\ov{\et}_x^a}+
\de^{da}\de(z-x)\imath q_{\ba x}\right]
\nonumber\\&&\left.
+\imath T_{\ba\al}^a\left[\frac{\de}{\de\imath c_z^d}\ev{\imath
\chi_{\ba x}\imath\ov{\et}_x^a}-\de^{da}\de(z-x)\imath\ov{q}
_{\ba x}\right]\ev{\imath\ov{q}_{\al x}}\right\}.
\label{eq:stid0}
\eea
Two further functional derivatives with respect to 
$\imath q_{\ga\w}$ and $\imath\ov{q}_{\de v}$ are taken and all 
remaing fields/sources set to zero.  Given that 
\cite{Watson:2008fb}
\be
\left.\frac{\de}{\de\imath c_z^d}\ev{\imath\ro_x^b\imath\ov{\et}_
x^c}\right|_{J=0}=\left.t_{ij}(\vec{x})\frac{\de}{\de\imath c_z^d}
\ev{\imath J_{jx}^b\imath\ov{\et}_x^c}\right|_{J=0}=0,
\ee
one obtains the \ST identity for the quark-gluon vertices in 
configuration space:
\bea
0&=\int d^4x\de(t-x_0)&\left\{
-\frac{\imath}{g}\left(\partial_x^0\ev{\imath\ov{q}_{\de v}\imath 
q_{\ga\w}\imath\si_x^d}\right)\de(z-x)
+\frac{1}{g}\left[\frac{\nabla_{ix}}{(-\nabla_x^2)}\ev{\imath
\ov{q}_{\de v}\imath q_{\ga\w}\imath A_{ix}^a}\right]\ev{\imath
\ov{c}_x^a\imath c_z^d}
\right.\nonumber\\&&
+\imath T_{\al\ba}^a\ev{\imath\ov{q}_{\de v}\imath q_{\al x}}
\left[\left.\frac{\de^2}{\de\imath q_{\ga\w}\de\imath c_z^d}
\ev{\imath\ov{\chi}_{\ba x}\imath\ov{\et}_x^a}\right|_{J=0}+
\de^{da}\de(z-x)\de_{\ga\ba}\de(\w-x)\right]
\nonumber\\&&\left.
+\imath T_{\ba\al}^a\left[\left.\frac{\de^2}{\de\imath\ov{q}
_{\de v}\de\imath c_z^d}\ev{\imath\chi_{\ba x}\imath\ov{\et}_x^a}
\right|_{J=0}-\de^{da}\de(z-x)\de_{\de\ba}\de(v-x)\right]\ev{\imath
\ov{q}_{\al x}\imath q_{\ga\w}}\right\}.
\label{eq:stid1}
\eea
As with all \ST identities, the above involves nontrivial ghost 
scattering-like kernels.  Introducing some notation, these can be 
expressed via the Legendre transform in terms of a loop integral 
over proper Green's functions and propagators in the following way:
\bea
\tilde{\G}_{\ov{q};\ov{c}cq\al\ga}^d(x,z,\w)&\equiv&\imath gT_{\al
\ba}^a\left.\frac{\de^2}{\de\imath q_{\ga\w}\de\imath c_z^d}\ev{
\imath\ov{\chi}_{\ba x}\imath\ov{\et}_x^a}\right|_{J=0}\nonumber\\
&=&\imath gT_{\al\ba}^a\left\{
-\ev{\imath\ov{\chi}_{\ba x}\imath\chi_\e}\ev{\imath\ov{q}_\e
\imath q_{\ga\w}\imath\Phi_\la}\ev{J_{\la}\imath J_\ka}\ev{\ov{\et}
_x^a\imath\et_\ta}\ev{\imath\ov{c}_\ta\imath c_z^d\imath\Phi_\ka}
\right.\nonumber\\&&\left.
+\ev{\imath\ov{\chi}_{\ba x}\imath\chi_\ka}\ev{\imath\ov{q}_\ka
\imath q_{\ga\w}\imath\ov{c}_\ta\imath c_z^d}\ev{\imath\ov{\et}_x^a
\imath\et_\ta}
\right\},\nonumber\\
\tilde{\G}_{q;\ov{c}c\ov{q}\de\al}^d(x,z,v)&\equiv&\left.
\frac{\de^2}{\de\imath\ov{q}_{\de v}\de\imath c_z^d}\ev{\imath\chi
_{\ba x}\imath\ov{\et}_x^a}\right|_{J=0}\imath gT_{\ba\al}^a
\nonumber\\
&=&\left\{
\ev{\ov{\et}_x^a\imath\et_\ta}\ev{\imath\ov{c}_\ta\imath c_z^d
\imath\Phi_\ka}\ev{\imath J_\ka\imath J_\la}\ev{\imath\ov{q}_
{\de v}\imath q_\e\imath\Phi_\la}\ev{\imath\ov{\chi}_\e\imath\chi_
{\ba x}}\right.\nonumber\\&&\left.
-\ev{\imath\ov{\et}_x^a\imath\et_\ta}\ev{\imath\ov{c}_\ta\imath 
c_z^d\imath\ov{q}_{\de v}\imath q_\ka}\ev{\imath\ov{\chi}_\ka\imath
\chi_{\ba x}}\right\}\imath gT_{\ba\al}^a.
\eea
In the above, $J$ and $\Phi$ are dummy sources/fields refering to 
either $\vec{A}$ or $\si$ only, the internal indices otherwise 
refer to all attributes of the object in question (summed or 
integrated over).  The configuration space \ST identity can thus 
be written
\bea
0&=\int d^4x\de(t-x_0)&\left\{
-\left(\imath\partial_x^0\ev{\imath\ov{q}_{\de v}\imath q_{\ga\w}
\imath\si_x^d}\right)\de(z-x)
+\left[\frac{\nabla_{ix}}{(-\nabla_x^2)}\ev{\imath\ov{q}_{\de v}
\imath q_{\ga\w}\imath A_{ix}^a}\right]\ev{\imath\ov{c}_x^a\imath 
c_z^d}\right.\nonumber\\&&\left.
+\ev{\imath\ov{q}_{\de v}\imath q_{\al x}}\left[\tilde{\G}_{\ov{q};
\ov{c}cq\al\ga}^d(x,z,\w)+\imath gT_{\al\ga}^d\de(z-x)\de(\w-x)
\right]\right.\nonumber\\&&\left.
+\left[\tilde{\G}_{q;\ov{c}c\ov{q}\de\al}^d(x,z,v)-\imath gT_{\de
\al}^d\de(z-x)\de(v-x)\right]\ev{\imath\ov{q}_{\al x}\imath q_{\ga
\w}}\right\}.
\label{eq:stid2}
\eea
Defining the Fourier transform for the vertex functions (all 
momenta incoming):
\be
\G(x,y,z)=\int\dk{k_1}\dk{k_2}\dk{k_3}(2\pi)^4\de(k_1+k_2+k_3)e^
{-\imath k_1\cdot x-\imath k_2\cdot y-\imath k_3\cdot z}\G(k_1,k_2,
k_3),
\ee
one can write the \ST identity in momentum space (dropping the 
dirac and fundamental color indices for convenience)
\bea
\lefteqn{
k_3^0\G_{\ov{q}q\si}^d(k_1,k_2,k_3)=\imath\frac{k_{3i}}{\vec{k}_
3^2}\G_{\ov{q}qAi}^a(k_1,k_2,k_3)\G_{\ov{c}c}^{ad}(-k_3)}
\nonumber\\&&
+\G_{\ov{q}q}(k_1)\left[\tilde{\G}_{\ov{q};\ov{c}cq}^d(k_1+q_0,k_3
-q_0,k_2)+\imath gT^d\right]+\left[\tilde{\G}_{q;\ov{c}c\ov{q}}^d
(k_2+q_0,k_3-q_0,k_1)-\imath gT^d\right]\G_{\ov{q}q}(-k_2).
\label{eq:stid3}
\eea
In the above, $k_1+k_2+k_3=0$, $\G_{\ov{c}c}$ is the proper ghost 
two-point function and $q_0$ is the (arbitrary) energy injection 
scale that arises from the time-dependence of the Gauss-BRST 
transform.  One can see the strong similarities of this expression 
with the \YM case \cite{Watson:2008fb}.  In momentum space, the 
kernels can be written
\bea
\tilde{\G}_{\ov{q};\ov{c}cq}^d(p_1,p_2,p_3)&=&\imath gT^a\int\dk
{\w}W_{\ov{c}c}^{ab}(\w)W_{\ov{q}q}(p_1-\w)
\nonumber\\&&\times
\left[\G_{\ov{c}c\ov{q}q}^{bd}(\w,p_2,p_1-\w,p_3)-\G_{\ov{c}c\ka}^
{bdc}(\w,p_2,-p_2-\w)W_{\ka\la}^{ce}(p_2+\w)\G_{\ov{q}q\la}^e(p_1-
\w,p_3,p_2+\w)\right],
\nonumber\\
\tilde{\G}_{q;\ov{c}c\ov{q}}^d(p_1,p_2,p_3)&=&\int\dk{\w}\left[\G_
{\ov{c}c\ka}^{bdc}(\w,p_2,-p_2-\w)W_{\ka\la}^{ce}(p_2+\w)\G_{\ov{q}
q\la}^{e}(p_3,p_1-\w,p_2+\w)-\G_{\ov{c}c\ov{q}q}^{bd}(\w,p_2,p_3,
p_1-\w)\right]
\nonumber\\&&\times
W_{\ov{c}c}^{ab}(\w)W_{\ov{q}q}(\w-p_1)\imath gT^a,
\label{eq:kern0}
\eea
where the indices $\ka$, $\la$, refer to the gluonic field types 
$\si$ or $\vec{A}$ (with the associated spatial index).  Just as 
in the \YM case, the above \ST identity, \eq{eq:stid3}, in 
conjunction with the kernels can be solved in principle to give 
the temporal quark-gluon vertex $\G_{\ov{q}q\si}$ in terms of 
purely spatial, ghost or quark propagators and proper functions.  
The identity is trivially satisfied at tree-level (using the 
Feynman rules of Ref.~\cite{Popovici:2008ty}) and can be verified 
at one-loop perturbatively (it is a purely technical exercise, so 
not suitable for presentation here).


\begin{thebibliography}{99}

%\cite{Sommer:1993ce}
\bibitem{Sommer:1993ce}
  R.~Sommer,
  %``A New way to set the energy scale in lattice gauge theories 
%and its  applications to the static force and alpha-s in SU(2) 
%Yang-Mills theory,''
  Nucl.\ Phys.\  B {\bf 411}, 839 (1994)
  [arXiv:hep-lat/9310022].
  %%CITATION = NUPHA,B411,839;%%

%\cite{Pak:2009em}
\bibitem{Pak:2009em}
  M.~Pak and H.~Reinhardt,
  %``The Wilson loop from a Dyson equation,''
  Phys.\ Rev.\  D {\bf 80}, 125022 (2009)
  [arXiv:0910.2916 [hep-th]].
  %%CITATION = PHRVA,D80,125022;%%

%\cite{Neubert:1993mb}
\bibitem{Neubert:1993mb}
  M.~Neubert,
  %``Heavy quark symmetry,''
  Phys.\ Rept.\  {\bf 245} (1994) 259
  [arXiv:hep-ph/9306320].
  %%CITATION = PRPLC,245,259;%%

%\cite{Mannel:1992fx}
\bibitem{Mannel:1992fx}
  T.~Mannel,
  %``Effective theory approach to systems with one heavy quark,''
  Chin.\ J.\ Phys.\  {\bf 31} (1993) 1.
  %%CITATION = CJOPA,31,1;%%

%\cite{Grinstein:1991ap}
\bibitem{Grinstein:1991ap}
  B.~Grinstein,
  ``Lectures on heavy quark effective theory,'' 
Mexico City HE Phenom.1991:0161-217.
  %%CITATION = C91-07-01.1;%%

%\cite{Eichten:1980mw}
\bibitem{Eichten:1980mw}
  E.~Eichten and F.~Feinberg,
  %``Spin Dependent Forces In QCD,''
  Phys.\ Rev.\  D {\bf 23}, 2724 (1981).
  %%CITATION = PHRVA,D23,2724;%%

%\cite{Alkofer:2002bp}
\bibitem{Alkofer:2002bp}
  R.~Alkofer, P.~Watson and H.~Weigel,
  %``Mesons in a Poincare covariant Bethe-Salpeter approach,''
  Phys.\ Rev.\  D {\bf 65}, 094026 (2002)
  [arXiv:hep-ph/0202053].
  %%CITATION = PHRVA,D65,094026;%%

%\cite{Maris:1999nt}
\bibitem{Maris:1999nt}
  P.~Maris and P.~C.~Tandy,
  %``Bethe-Salpeter study of vector meson masses and decay 
%constants,''
  Phys.\ Rev.\  C {\bf 60}, 055214 (1999)
  [arXiv:nucl-th/9905056].
  %%CITATION = PHRVA,C60,055214;%%

%\cite{Fischer:2006ub}
\bibitem{Fischer:2006ub}
  C.~S.~Fischer,
  %``Infrared properties of QCD from Dyson-Schwinger equations,''
  J.\ Phys.\ G {\bf 32}, R253 (2006)
  [arXiv:hep-ph/0605173].
  %%CITATION = JPHGB,G32,R253;%%

%\cite{Watson:2004kd}
\bibitem{Watson:2004kd}
  P.~Watson, W.~Cassing and P.~C.~Tandy,
  %``Bethe-Salpeter meson masses beyond ladder approximation,''
  Few Body Syst.\  {\bf 35}, 129 (2004)
  [arXiv:hep-ph/0406340].
  %%CITATION = FBSYE,35,129;%%

%\cite{Williams:2009wx}
\bibitem{Williams:2009wx}
  R.~Williams,
  %``Bethe-Salpeter studies of mesons beyond rainbow-ladder,''
  arXiv:0912.3494 [hep-ph].
  %%CITATION = ARXIV:0912.3494;%%

%\cite{Bender:2002as}
\bibitem{Bender:2002as}
  A.~Bender, W.~Detmold, C.~D.~Roberts and A.~W.~Thomas,
  %``Bethe-Salpeter equation and a nonperturbative quark gluon 
%vertex,''
  Phys.\ Rev.\  C {\bf 65}, 065203 (2002)
  [arXiv:nucl-th/0202082].
  %%CITATION = PHRVA,C65,065203;%%

%\cite{Bhagwat:2004hn}
\bibitem{Bhagwat:2004hn}
  M.~S.~Bhagwat, A.~Holl, A.~Krassnigg, C.~D.~Roberts and 
P.~C.~Tandy,
  %``Aspects and consequences of a dressed-quark-gluon vertex,''
  Phys.\ Rev.\  C {\bf 70}, 035205 (2004)
  [arXiv:nucl-th/0403012].
  %%CITATION = PHRVA,C70,035205;%%

%\cite{Matevosyan:2006bk}
\bibitem{Matevosyan:2006bk}
  H.~H.~Matevosyan, A.~W.~Thomas and P.~C.~Tandy,
  %``Quark-gluon vertex dressing and meson masses beyond 
%ladder-rainbow  truncation,''
  Phys.\ Rev.\  C {\bf 75}, 045201 (2007)
  [arXiv:nucl-th/0605057].
  %%CITATION = PHRVA,C75,045201;%%

%\cite{Bender:1996bb}
\bibitem{Bender:1996bb}
  A.~Bender, C.~D.~Roberts and L.~Von Smekal,
  %``Goldstone Theorem and Diquark Confinement Beyond 
%Rainbow-Ladder  Approximation,''
  Phys.\ Lett.\  B {\bf 380}, 7 (1996)
  [arXiv:nucl-th/9602012].
  %%CITATION = PHLTA,B380,7;%%

%\cite{Fischer:2005en}
\bibitem{Fischer:2005en}
  C.~S.~Fischer, P.~Watson and W.~Cassing,
  %``Probing unquenching effects in the gluon polarisation in 
%light mesons,''
  Phys.\ Rev.\  D {\bf 72}, 094025 (2005)
  [arXiv:hep-ph/0509213].
  %%CITATION = PHRVA,D72,094025;%%

%\cite{Watson:2004jq}
\bibitem{Watson:2004jq}
  P.~Watson and W.~Cassing,
  %``Unquenching the quark antiquark Green's function,''
  Few Body Syst.\  {\bf 35}, 99 (2004)
  [arXiv:hep-ph/0405287].
  %%CITATION = FBSYE,35,99;%%

%\cite{Fischer:2009jm}
\bibitem{Fischer:2009jm}
  C.~S.~Fischer and R.~Williams,
  %``Probing the gluon self-interaction in light mesons,''
  Phys.\ Rev.\ Lett.\  {\bf 103}, 122001 (2009)
  [arXiv:0905.2291 [hep-ph]].
  %%CITATION = PRLTA,103,122001;%%

%\cite{Adler:1984ri}
\bibitem{Adler:1984ri}
  S.~L.~Adler and A.~C.~Davis,
  %``CHIRAL SYMMETRY BREAKING IN COULOMB GAUGE QCD,''
  Nucl.\ Phys.\  B {\bf 244}, 469 (1984).
  %%CITATION = NUPHA,B244,469;%%

%\cite{Alkofer:2005ug}
\bibitem{Alkofer:2005ug}
  R.~Alkofer, M.~Kloker, A.~Krassnigg and R.~F.~Wagenbrunn,
  %``Aspects of the confinement mechanism in Coulomb-gauge QCD,''
  Phys.\ Rev.\ Lett.\  {\bf 96}, 022001 (2006)
  [arXiv:hep-ph/0510028].
  %%CITATION = PRLTA,96,022001;%%

%\cite{Abers:1973qs}
\bibitem{Abers:1973qs}
  E.~S.~Abers and B.~W.~Lee,
  %``Gauge Theories,''
  Phys.\ Rept.\  {\bf 9}, 1 (1973).
  %%CITATION = PRPLC,9,1;%%

%\cite{Gribov:1977wm}
\bibitem{Gribov:1977wm}
  V.~N.~Gribov,
  %``Quantization of non-Abelian gauge theories,''
  Nucl.\ Phys.\ B {\bf 139} (1978) 1.
  %%CITATION = NUPHA,B139,1;%%

%\cite{Zwanziger:1995cv}
\bibitem{Zwanziger:1995cv}
  D.~Zwanziger,
  %``Lattice Coulomb Hamiltonian and static color-Coulomb field,''
  Nucl.\ Phys.\  B {\bf 485}, 185 (1997)
  [arXiv:hep-th/9603203].
  %%CITATION = NUPHA,B485,185;%%

%\cite{Zwanziger:1998ez}
\bibitem{Zwanziger:1998ez}
  D.~Zwanziger,
  %``Renormalization in the Coulomb gauge and order parameter for 
  % confinement in QCD,''
  Nucl.\ Phys.\ B {\bf 518} (1998) 237.
  %%CITATION = NUPHA,B518,237;%%

%\cite{Reinhardt:2008pr}
\bibitem{Reinhardt:2008pr}
  H.~Reinhardt and P.~Watson,
  %``Resolving temporal Gribov copies in Coulomb gauge Yang-Mills 
%theory,''
  arXiv:0808.2436 [hep-th].
  %%CITATION = ARXIV:0808.2436;%%

%\cite{Watson:2006yq}
\bibitem{Watson:2006yq}
  P.~Watson and H.~Reinhardt,
  %``Propagator Dyson-Schwinger equations of Coulomb gauge 
%Yang-Mills theory  within the first order formalism,''
  Phys.\ Rev.\  D {\bf 75}, 045021 (2007)
  [arXiv:hep-th/0612114].
  %%CITATION = PHRVA,D75,045021;%%

%\cite{Watson:2007vc}
\bibitem{Watson:2007vc}
  P.~Watson and H.~Reinhardt,
  %``Two-Point Functions of Coulomb Gauge Yang-Mills Theory,''
  Phys.\ Rev.\  D {\bf 77}, 025030 (2008)
  [arXiv:0709.3963 [hep-th]].
  %%CITATION = PHRVA,D77,025030;%%

%\cite{Popovici:2008ty}
\bibitem{Popovici:2008ty}
  C.~Popovici, P.~Watson and H.~Reinhardt,
  %``Quarks in Coulomb gauge perturbation theory,''
  Phys.\ Rev.\  D {\bf 79}, 045006 (2009)
  [arXiv:0810.4887 [hep-th]].
  %%CITATION = PHRVA,D79,045006;%%

%\cite{Watson:2008fb}
\bibitem{Watson:2008fb}
  P.~Watson and H.~Reinhardt,
 %``Slavnov-Taylor identities in Coulomb gauge Yang-Mills theory,''
  Eur.\ Phys.\ J.\  C {\bf 65} (2010) 567
  [arXiv:0812.1989 [hep-th]].
  %%CITATION = EPHJA,C65,567;%%

%\cite{Watson:2007mz}
\bibitem{Watson:2007mz}
  P.~Watson and H.~Reinhardt,
  %``Perturbation Theory of Coulomb Gauge Yang-Mills Theory 
%Within the First  Order Formalism,''
  Phys.\ Rev.\  D {\bf 76}, 125016 (2007)
  [arXiv:0709.0140 [hep-th]].
  %%CITATION = PHRVA,D76,125016;%%

%\cite{Burgio:2008jr}
\bibitem{Burgio:2008jr}
  G.~Burgio, M.~Quandt and H.~Reinhardt,
  %``Coulomb gauge gluon propagator and the Gribov formula,''
  Phys.\ Rev.\ Lett.\  {\bf 102} (2009) 032002
  [arXiv:0807.3291 [hep-lat]].
  %%CITATION = PRLTA,102,032002;%%

%\cite{Quandt:2008zj}
\bibitem{Quandt:2008zj}
  M.~Quandt, G.~Burgio, S.~Chimchinda and H.~Reinhardt,
  %``Coulomb gauge ghost propagator and the Coulomb potential,''
  PoS {\bf CONFINEMENT8}, 066 (2008)
  [arXiv:0812.3842 [hep-th]].
  %%CITATION = POSCI,CONFINEMENT8,066;%%

%\cite{Feuchter:2004mk}
\bibitem{Feuchter:2004mk}
  C.~Feuchter and H.~Reinhardt,
  %``Variational solution of the Yang-Mills Schroedinger 
%equation in Coulomb  gauge,''
  Phys.\ Rev.\ D {\bf 70} (2004) 105021
  [arXiv:hep-th/0408236];
  %%CITATION = HEP-TH 0408236;%%
  C.~Feuchter and H.~Reinhardt,
  %``Quark and gluon confinement in Coulomb gauge,''
  arXiv:hep-th/0402106.
  %%CITATION = HEP-TH/0402106;%%

%\cite{Epple:2006hv}
\bibitem{Epple:2006hv}
  D.~Epple, H.~Reinhardt and W.~Schleifenbaum,
  %``Confining Solution of the Dyson-Schwinger Equations 
%in Coulomb Gauge,''
  Phys.\ Rev.\  D {\bf 75}, 045011 (2007)
  [arXiv:hep-th/0612241].
  %%CITATION = PHRVA,D75,045011;%%

%\cite{Epple:2007ut}
\bibitem{Epple:2007ut}
  D.~Epple, H.~Reinhardt, W.~Schleifenbaum and A.~P.~Szczepaniak,
  %``Subcritical solution of the Yang-Mills Schroedinger
% equation in the Coulomb  gauge,''
  Phys.\ Rev.\  D {\bf 77} (2008) 085007
  [arXiv:0712.3694 [hep-th]].
  %%CITATION = PHRVA,D77,085007;%%

%\cite{Szczepaniak:2001rg}
\bibitem{Szczepaniak:2001rg}
  A.~P.~Szczepaniak and E.~S.~Swanson,
  %``Coulomb gauge QCD, confinement, and the constituent 
%representation,''
  Phys.\ Rev.\  D {\bf 65}, 025012 (2002)
  [arXiv:hep-ph/0107078];
  %%CITATION = PHRVA,D65,025012;%%
%\cite{Szczepaniak:2003ve}
%\bibitem{Szczepaniak:2003ve}
  A.~P.~Szczepaniak,
  %``Confinement and gluon propagator in Coulomb gauge QCD,''
  Phys.\ Rev.\  D {\bf 69}, 074031 (2004)
  [arXiv:hep-ph/0306030].
  %%CITATION = PHRVA,D69,074031;%%

%\cite{Cucchieri:2000hv}
\bibitem{Cucchieri:2000hv}
  A.~Cucchieri and D.~Zwanziger,
 %``Renormalization group calculation of color Coulomb potential,''
  Phys.\ Rev.\  D {\bf 65}, 014002 (2001)
  [arXiv:hep-th/0008248].
  %%CITATION = PHRVA,D65,014002;%%

%\cite{Ball:1980ay}
\bibitem{Ball:1980ay}
  J.~S.~Ball and T.~W.~Chiu,
  %``Analytic Properties Of The Vertex Function In Gauge Theories.
% 1,''
  Phys.\ Rev.\  D {\bf 22}, 2542 (1980).
  %%CITATION = PHRVA,D22,2542;%%

%\cite{Zwanziger:2002sh}
\bibitem{Zwanziger:2002sh}
  D.~Zwanziger,
  %``No confinement without Coulomb confinement,''
  Phys.\ Rev.\ Lett.\  {\bf 90}, 102001 (2003)
  [arXiv:hep-lat/0209105].
  %%CITATION = PRLTA,90,102001;%%

\end{thebibliography}
\end{document}